\documentclass[12pt,preprint]{aastex}

\begin{document}

\title{Quasi-Periodic Oscillations from Rayleigh-Taylor and
Kelvin-Helmholtz Instability at a Disk-Magnetosphere Interface}

\author{Li-Xin Li\altaffilmark{1}\, and Ramesh Narayan}
\affil{Harvard-Smithsonian Center for Astrophysics, Cambridge,
MA 02138, USA} \email{lli,rnarayan@cfa.harvard.edu}

\altaffiltext{1}{Chandra Fellow}

\begin{abstract}

We consider the interface between an accretion disk and a 
magnetosphere surrounding the accreting mass.  We argue that such an
interface can occur not only with a magnetized neutron star but also
sometimes with an unmagnetized neutron star or a black hole.  The gas
at the magnetospheric interface is generally Rayleigh-Taylor unstable
and may also be Kelvin-Helmholtz unstable.  Because of these
instabilities, modes with low azimuthal wavenumbers $m$ are expected
to grow to large amplitude.  It is proposed that the resulting
nonaxisymmetric structures contribute to the high frequency
quasi-periodic oscillations that have been seen in neutron-star and
black-hole X-ray binaries.  The mode oscillation frequencies are
calculated to be approximately equal to $m \Omega_m$, where $\Omega_m$ is the
angular velocity of the accreting gas at the magnetospheric radius.
Thus, mode frequencies should often be in the approximate ratio 1:2:3,
etc.  If the pressure of the gas in the disk is not large, then the
$m=1$ mode will be stable.  In this case, the mode frequencies should
be in the approximate ratio 2:3, etc.  There is some observational
evidence for such simple frequency ratios.

\end{abstract}


\section{Introduction}
\label{sec1}

The study of the timing properties of X-ray binaries and, in
particular, quasi-periodic oscillations (QPOs) in these sources, has
for long been a major area of research (see van der Klis 2000 and
Remillard et al. 2002b for recent reviews, and Lewin, van Paradijs, \&
van der Klis 1988 for a discussion of earlier observations).  The
field has become especially active in recent years after the launch of
the {\it Rossi X-Ray Timing Explorer} \citep[{\it RXTE};][]{bra93}
with its superb timing capability.  Among the divergent phenomenology
of QPOs observed in X-ray binaries, of particular interest are the
``kilohertz QPOs,'' which have frequencies $\sim 10^2-10^3$Hz.

About twenty accreting neutron-star sources and five accreting
black-hole sources are known to exhibit kHz QPOs
\citep{kli00,rem02}. These QPOs often appear in pairs (so called
``twin QPOs''), with the frequency difference being anticorrelated to
the QPO frequencies \citep[see, however, Migliari, van der Klis, \& 
Fender 2003]{kli00,sal03}.  In many neutron star systems
the frequency difference appears to be related to the spin frequency
of the neutron star \citep{kli00,wij03}.  At the
same time, in several sources, the twin kHz QPO frequencies seem to
occur in the ratio of simple integers, with a frequency ratio of 2:3
being common in both black hole and neutron star systems
\citep{abr01,abr03,rem02a}.

Kilohertz QPOs occur at a similar frequency in neutron star sources
that differ in X-ray luminosity by more than two orders of magnitude.
Moreover, the QPO frequency in a given source seems to track the
fluctuation in the X-ray intensity from average for that source rather
than the absolute luminosity of the source
\citep{kli97,kli98,kli00,zha97}.  Furthermore, there is a surprising
continuity of QPO properties between black hole and neutron star
binaries (Psaltis, Belloni \& van der Klis 1999; Belloni, Psaltis \&
van der Klis 2002).  The continuity even appears to extend to white
dwarfs (Warner, Woudt \& Pretorius 2003).

Because of their high frequencies, kHz QPOs must be produced by
processes close to the accreting mass.  However, since these QPOs have
been observed in both neutron-star and black-hole X-ray binaries, the
oscillations are unlikely to be associated with the surface of the
accreting object.  Instead, it seems likely that the kilohertz QPOs
originate in the accretion flow surrounding the central mass.
Motivated by this argument, a variety of accretion-based models have
been proposed to explain the oscillations \citep[and references
therein]{lew88,kli00}.

The classical beat frequency model for QPOs in neutron-star sources
proposes that one of the kHz QPOs is associated with the Keplerian
frequency of the orbiting gas at the inner edge of the accretion disk
and that the second kHz QPO represents a beat phenomenon between the
orbiting gas and the spin of the central star
\citep{alp85,lam85,str96,mil98,lam01}.  This model predicts that the
frequency difference between the two kHz QPOs should be equal to the
spin frequency of the neutron star and should be constant. However,
observations show that the separation is not a constant: it generally
decreases when the QPO frequencies increase.  The frequency difference
is also observed not to be precisely equal to the frequency of burst
oscillations \citep{kli00}, which are believed to match the spin
frequency of the star. Most interestingly, kHz QPOs have recently been
detected in an accreting millisecond pulsar, SAX J1808.4--3658, whose
spin frequency is known \citep{wij03}. The $\sim 195$-Hz frequency
difference between the two kHz QPOs in this source is far below the
$401$-Hz spin frequency of the neutron star, but is consistent with
half the spin frequency.  \citet{wij03} argue that their observations
falsify the beat-frequency model and pose a severe challenge for all
current models of the kHz QPOs (but see Lamb \& Miller 2003).

Stella \& Vietri (1998,1999) proposed that the QPOs may be interpreted
in terms of fundamental frequencies of test particles in motion in the
relativistic potential of a neutron star or a black hole.  They
identified individual QPO frequencies with the orbital, periastron
precession and nodal precession frequencies of test particles and
showed that some predicted scalings between the frequencies agree with
observations (Stella \& Vietri 1999; Stella, Vietri \& Morsink 1999).
In contrast to some versions of the beat frequency model, this model
does not require a magnetic field anchored in the central star, and
thus provides a natural explanation for why twin QPOs are seen in both
neutron-star and black-hole systems.  However, the fact that the
frequency difference between twin kHz QPOs in neutron star systems is
often of order the neutron star frequency has no simple explanation.
Also, since the model explicitly invokes general relativistic effects,
it cannot explain the continuity in QPO properties between neutron
stars and black holes on the one hand and white dwarfs on the other
(Warner et al. 2003).

A number of scientists have investigated modes of oscillation of an
accretion disk as a model of high frequency QPOs \citep[for reviews
see][]{kat98,wag99,kat01a}. In an important paper, \citet{oka87}
showed that axisymmetric $g$-modes are trapped in the inner regions of
a relativistic disk, where the epicyclic frequency $\kappa$ reaches a
maximum. This idea was exploited by \citet{now91,now92} and a number
of other workers \citep{per97,now97,sil01,wag01,abr01} who worked out 
the physical
properties of the trapped modes.  Much of the work has focused on
neutral (i.e., non-growing) modes.  The idea in such models is that,
although the modes are stable, they might nevertheless be excited by
some driving mechanism such as disk turbulence to produce the observed
QPOs.

Recently, \citet{kat01b,kat02} claimed to find that nonaxisymmetric
$g$-modes are trapped between two forbidden zones that lie on either
side of the corotation radius and that the modes are highly unstable.
Such dynamically unstable modes are very interesting since they would
be self-excited and would spontaneously grow to non-linear amplitudes
without the need for an external driving mechanism.  Motivated by
Kato's work, \citet{li03} studied nonaxisymmetric $g$-mode and
$p$-mode instabilities in an unmagnetized isothermal accretion disk.
They found that $g$- and $p$-modes with a nonzero number of vertical
nodes are strongly absorbed at corotation and thus not
amplified.  Waves without vertical nodes are amplified
(as known earlier, see Papaloizou \& Pringle 1985; Goldreich, Goodman,
\& Narayan 1986; Narayan, Goldreich, \& Goodman 1987), but the
amplification is very weak.  Therefore, any energy loss, either during
propagation of the wave or during reflection at the boundaries, would
kill the instability. Thus, Li et al. (2003) concluded, in agreement
with Kato (2003), that nonaxisymmetric disk modes are not
sufficiently unstable to be of interest for the QPO problem.
Ortega-Rodriguez \& Wagoner (2000) found that viscosity causes the
fundamental $g$- and $p$-modes in a disk to grow, though the relevance of
their result for QPOs is presently unclear.

Since the amplitude of intensity fluctuations observed in QPOs is
often quite large, it is the opinion of the present authors that the
oscillations are likely to be the result of a strongly-growing {\it
instability} of some sort in the accretion flow.  Having argued above
that a hydrodynamic instability is ruled out (Kato 2003; Li et
al. 2003), it is natural to consider the effect of magnetic fields via
a magnetohydrodynamic (MHD) instability.  Since, in general, only a 
small number of
frequencies dominate the intensity fluctuations, the instability
cannot be present at all disk radii (as in the case of the
magnetorotational instability, Balbus \& Hawley 1998), but must be
associated with some special radius in the disk.  The most natural
choice for the special radius is the inner edge of the disk.
Motivated by these arguments, we consider in this paper an accretion
disk that is terminated at its inner edge by a strong vertical
magnetic field.  The resulting magnetospheric interface suffers from
both the Rayleigh-Taylor and Kelvin-Helmholtz instabilities.  We
calculate the frequencies and growth rates of the unstable modes and
consider the role that these modes play in the kHz QPOs seen in X-ray
binaries.

The Rayleigh-Taylor instability and the related interchange
instability have been considered by a number of authors, both for
modeling QPOs (Titarchuk 2002, 2003) and in connection with other
applications (Arons \& Lea 1976a,b; Elsner \& Lamb 1976, 1977;
Michel 1977a,b; Spruit \& Taam 1990; Kaisig, Tajima \& Lovelace 1992;
Lubow \& Spruit 1995; Spruit, Stehle \& Papaloizou 1995; Chandran
2001).  We compare our work to these earlier studies.

The plan of the paper is as follows.  In \S2 we describe the model and
its initial equilibrium state. In \S3 we consider linear
perturbations, derive a wave equation satisfied by the perturbations,
and obtain the corresponding jump conditions across the magnetospheric
radius where there is a discontinuity in disk properties. In \S4 we
consider the case when the mass density of the disk has a jump but the
angular velocity is continuous across the boundary; this situation
occurs when the central object is a black hole. We show that, under
suitable conditions, there is a Rayleigh-Taylor instability at the
boundary, and we study in detail the growth of the instability using
both analytical and numerical methods. In \S5 we study the case when
both the mass density and the angular velocity of the disk are
discontinuous at the boundary, which corresponds to the situation when
the central object is a neutron star. We show that both the
Rayleigh-Taylor and the Kelvin-Helmholtz instabilities are present
under suitable conditions. In \S6 we summarize and discuss the
results, and briefly compare the predictions of the model to
observations of QPOs. In Appendix~A we derive a necessary condition
for disk instability.  In Appendix~B we briefly review classical
results on the Rayleigh-Taylor and Kelvin-Helmholtz instabilities in
Cartesian flows.

\section{Basic Model and Initial Equilibrium}
\label{sec2}

\subsection{Outline of the Model}
\label{sec2.1}

We consider a differentially-rotating axisymmetric accretion flow that
is terminated on the inside by a strong magnetic field.  The interface
between the disk and the magnetosphere occurs at the magnetospheric
radius $r_m$. The density experiences a jump at $r_m$, being larger on
the outside and smaller on the inside.  There may also be a jump in
the angular velocity.  Because of the density and velocity jumps, the
system is potentially Rayleigh-Taylor and Kelvin-Helmholtz unstable
\citep{cha61,dra81}.

The geometry we envisage is natural for accretion onto a magnetized
neutron star. This geometry has been studied for many years (e.g.,
Ghosh \& Lamb 1978, 1979) and there are several discussions of
possible instabilities at the disk-magnetosphere interface
\citep{aro76a,aro76b,els76,els77,mic77a,mic77b,ikh96}. It has been
suggested that the interchange instability (i.e., the Rayleigh-Taylor
instability in the presence of magnetic fields) is the fastest mode of
mass transport into the magnetosphere of an accreting magnetized
neutron star \citep{els84}. When the accretion flow is spherical and
the neutron star rotates slowly, the instability criterion is
determined by a competition between the gravitational force of the
neutron star and the poloidal curvature force of the magnetic field:
gravity tends to drive the Rayleigh-Taylor instability, while the
curvature force tends to stabilize the configuration
\citep{aro76a,aro76b,els76,els77,mic77a,mic77b}. When the neutron star
rotates fast enough, the magnetospheric boundary becomes stable with
respect to the interchange instability because of the strong toroidal
magnetic field that is produced in the boundary layer by the rotation
of the magnetosphere relative to the spherical accretion flow
\citep{ikh96}.

Similar processes for a thin Keplerian disk accreting onto a
magnetized neutron star have been analysed by \citet{spr90}. These
authors confirmed that the Rayleigh-Taylor instability can be very
effective in transporting mass across field lines, allowing the gas to
drift along the midplane in a disk-like fashion even in regions of the
magnetosphere that corotate with the central star.  The interchange
instability in a general thin accretion disk with differential
rotation has been studied by \citet{lub95} and \citet{spr95}. They
found that disk shear tends to stabilize the configuration.

The model we consider differs from the studies cited above in one
important respect: we explicitly avoid field curvature and focus on
the pure Rayleigh-Taylor and Kelvin-Helmholtz problem for a
disk-magnetosphere interface.  The model is thus highly simplified and
is perhaps most relevant for gas in the mid-plane of the accretion
disk. We also ignore the toroidal component of the magnetic field,
even though it may often be important (Ghosh \& Lamb 1978; Ikhsanov \&
Pustil'nik 1996). A toroidal magnetic field affects the stability of
the disk in two ways: (1) The toroidal field causes an inward force
because of its curvature, enhancing the effective gravity in the disk
and thereby boosting the Rayleigh-Taylor instability. (2) The toroidal
magnetic field also produces a tension at the interface which tends to
stabilize the disk (see, e.g, Chandrasekhar 1961). The overall effect
will be determined by the competition between these two tendencies. We
expect that for modes with a low azimuthal wave number the two effects
will approximately cancel each other so that, at least for modest
levels of toroidal field, the results may not be affected
significantly.  However, for modes with a large azimuthal wave number,
the second effect will be dominant and so a toroidal field will very
likely stabilize the disk.

Although our model is clearly relevant for neutron star accretion, it
is in fact relevant also for accretion onto a black hole or an
unmagnetized neutron star.  Bisnovatyi-Kogan \& Ruzmaikin (1974, 1976)
proposed that an accretion flow around a black hole may, under certain
circumstances, advect a considerable amount of magnetic flux to the
center.  Once enough magnetic flux has collected, the accumulated
field will disrupt the flow at a magnetospheric radius, exactly as in
our model.  This idea has been confirmed in recent three-dimensional
MHD simulations of radiatively inefficient accretion flows by
\citet{igu03}, and some additional consequences have been explored by
Narayan, Igumenshchev \& Abramowicz (2003).  Something like a
magnetosphere may also result from a global poloidal magnetic field
being generated within the disk (Livio, Pringle, \& King 2003).

The angular velocity profile has some important qualitative
differences between the black hole and neutron star problems.  In the
black hole case, we expect the angular velocity to be continuous
across $r_m$ (\S\ref{sec4}, Fig.~\ref{fig2}), whereas in the neutron
star case the angular velocity will in general be discontinuous at
$r_m$ (\S\ref{sec5}, Fig.~\ref{fig6}).  Thus, the disk-magnetosphere
interface in the case of a black hole experiences only the
Rayleigh-Taylor instability, whereas in the case of a neutron star it
can have both the Rayleigh-Taylor and Kelvin-Helmholtz instabilities.

In order to be able to handle both the black hole and neutron star
problems, we set up the theoretical framework sufficiently generally
as far as the angular velocity profile is concerned.  However, to make
the analysis tractable, we make some other drastic simplifications.
First, we assume that the gas is incompressible. With this
approximation, compressive waves are filtered out.  The interchange
and Kelvin-Helmholtz instabilities, which are our main interest and
which are both incompressible in nature, survive unaffected
\citep{spr95}.  Second, we assume that the magnetic field is purely
along the $z$-axis (i.e., parallel to the rotation axis of the gas
flow) and that all fluid quantities are independent of $z$.  We thus
solve a cylindrically symmetric problem in which $\partial/\partial
z=0$ and all quantities are function only of the cylindrical radius
$r$ and the azimuthal angle $\phi$.  The geometry of the model is as
shown in Figure~\ref{fig1}.

We describe below in \S\ref{sec2.2} the equilibrium flow, and analyse
in \S\ref{sec3} the effect of linear perturbations.

\subsection{Basic Equations and the Initial Equilibrium State}
\label{sec2.2}

The dynamics of a perfectly conducting and magnetized Newtonian
accretion disk is governed by the basic equations of MHD \citep[we
neglect viscosity and resistivity]{bal98}:
\begin{eqnarray}
     \frac{\partial\rho}{\partial t} + \nabla\cdot(\rho{\bf v})              
	     = 0 \;,\hspace{3.6cm}\label{cont_eq} \\
     \rho\left(\frac{\partial}{\partial t}+ {\bf v}\cdot\nabla
          \right){\bf v} = -\rho\nabla\Phi - \nabla\left(p +
          \frac{B^2}{8\pi}\right) + \frac{1}{4\pi} {\bf B}\cdot
          \nabla{\bf B} \;, \label{mom_eq} \\
     \frac{\partial{\bf B}}{\partial t} = \nabla\times({\bf v}\times
          {\bf B}) \;, \hspace{3.5cm}\label{free}\\
     \nabla\cdot{\bf B} = 0 \;, \hspace{4.4cm}\label{div} 
\end{eqnarray}
where $\rho$ is the mass density, ${\bf v}$ is the velocity, $\Phi$ is
the gravitational potential of the central compact object, $p$ is the
gas pressure, and ${\bf B}$ is the magnetic field.
Equations~(\ref{cont_eq})--(\ref{free}) are respectively the
continuity, momentum and induction equations, and equation~(\ref{div})
is one of Maxwell equations.

We assume that the disk fluid is incompressible, i.e.
\begin{eqnarray}
     \nabla\cdot{\bf v} = 0 \;.
     \label{divv}
\end{eqnarray}
Substituting this in the continuity equation~(\ref{cont_eq}) gives
\begin{eqnarray}
     \frac{d\rho}{dt} = 0 \;,  \hspace{1cm}
	\frac{d}{d t} \equiv \frac{\partial}{\partial t} + {\bf v}
          \cdot\nabla \;,
     \label{incom}
\end{eqnarray}
where $d/dt$ is the Lagrangian time derivative as measured in the
comoving frame of the fluid.  Equation~(\ref{incom}) states that the
mass density measured by a comoving observer does not change with
time, which is obvious for an incompressible fluid.

As explained in \S\ref{sec2.1}, the equilibrium state is assumed to be 
described by
\begin{eqnarray}
     \rho = \rho_0(r) \;, \qquad 
	p = p_0(r) \;, \qquad
	{\bf v} = \Omega(r) r \, {\bf e}_\phi \;, \qquad
	{\bf B} = B_0(r) \, {\bf e}_z \;,
	\label{v0etc}
\end{eqnarray}
where ${\bf e}_i$ ($i = r, \phi,z$) are unit coordinate vectors, and we use 
the subscript 0 to refer to the equilibrium. Equations~(\ref{cont_eq}),
(\ref{free})--(\ref{incom}) are automatically satisfied by such an equilibrium, 
while equation~(\ref{mom_eq}) gives an additional requirement:
\begin{eqnarray}
	g_{\rm eff} = -{1\over\rho_0}{dp_{t0}\over dr} \;.
	\label{equi_r}
\end{eqnarray}
Here, the effective gravitational acceleration $g_{\rm eff}$ and the
associated frequency $\Omega_{\rm eff}$ are defined by
\begin{eqnarray}
     g_{\rm eff} \equiv \Omega_{\rm eff}^2 r
	     = \frac{d\Phi}{dr} - \Omega^2 r \;,
     \label{geff}
\end{eqnarray}
and the total pressure $p_t$ is given by
\begin{eqnarray}
	p_t \equiv p + {B^2\over 8\pi} \;.
	\label{ptot}
\end{eqnarray}
Equation (\ref{equi_r}) states that the net radial force, obtained by
summing the gravitational, centrifugal and (total) pressure gradient
forces, vanishes in equilibrium.  Note that, to be consistent with the
assumption of cylindrical symmetry, the background gravitational
potential $\Phi$ must be a function of only the cylindrical radius
$r$.  This is one of the many simplifications we have made in the
model.

As explained in \S\ref{sec2.1}, the initial equilibrium consists of
two distinct zones separated at a radius $r_m$ (Fig. \ref{fig1}).  At
$r_m$, there is a discontinuity in $\rho$ and ${\bf B}$, and sometimes
also in $\Omega$ and ${\bf v}$ (see \S5 below).  However, the pressure
$p$ is always continuous.

\section{Linear Perturbations}
\label{sec3}

\subsection{Wave Equation}
\label{sec3.1}

We now consider linear perturbations of the equilibrium configuration.
We take the perturbations to be proportional to $\exp[i (-\omega t +
m\phi)]$, where $\omega$ is the mode frequency, which is in general
complex, and $m$ is the number of nodes in the $\phi$ direction; the
azimuthal wavevector is equal to $m/r$.  We assume that the
perturbations are independent of $z$.  Therefore, magnetic field lines
which are initially parallel to $z$ (by assumption) remain so even in
the perturbed state, and the $r$- and $\phi$-components of the
magnetic field vanish at all times.  We then have from
equation~({\ref{mom_eq}),
\begin{eqnarray}
     {\bf B}\cdot\nabla \propto \frac{\partial}{\partial z} = 0 \;.
	\label{bdot}
\end{eqnarray}
Since we have further assumed that the gravitational potential $\Phi$
is a function only of $r$, none of the force terms on the right-hand
side of equation~(\ref{mom_eq}) has a vertical component.  Therefore,
the vertical velocity perturbation is also zero.  The linear
perturbations to the mass density, gas pressure, velocity, and
magnetic field may be written as
\begin{eqnarray}
     \left\{\begin{array}{c}
	      \delta\rho \\ \delta p \\ \delta{\bf v} \\ \delta{\bf B}
	      \end{array}\right\}
	     = \left\{\begin{array}{c}
		 \rho_1(r) \\ p_1(r) \\ u(r)\, {\bf e}_r + v(r)\, {\bf e}_\phi
		 \\ B_1(r)\, {\bf e}_z\end{array}\right\}
		\times e^{-i\omega t + im\phi} \;. 
	     \label{drho}
\end{eqnarray}
Taking the curl of equation~(\ref{mom_eq}) and substituting
equations~(\ref{divv}) and (\ref{bdot}) gives the following equation
for the vorticity,
\begin{eqnarray}
     \frac{d}{dt}\left(\nabla\times{\bf v}\right) = -\frac{1}{\rho}
	  \nabla\rho\times\left(\nabla\Phi + \frac{d{\bf v}}{dt}\right) \;.
     \label{cmom_eq}
\end{eqnarray}

Equations~(\ref{divv}), (\ref{incom}), and (\ref{cmom_eq}) then form a
complete set of equations for ${\bf v}$ and $\rho$.  These equations
are independent of the gas pressure and the magnetic field.  The first
order perturbations of equations~(\ref{divv}) and (\ref{incom}) give
\begin{eqnarray}
     \frac{1}{r}\frac{d}{dr}(ru) + \frac{im}{r}\,v &=& 0 \;, 
     \label{pdivv} \\
	i\sigma\rho_1 - u\,\frac{d\rho_0}{dr} &=& 0\;, 
	\label{pincom}
\end{eqnarray}
respectively, where 
\begin{eqnarray}
	\sigma \equiv \omega - m \Omega \;.
	\label{sigma}
\end{eqnarray}
Similarly, the first order perturbation of equation~(\ref{cmom_eq})
gives
\begin{eqnarray}
     \frac{1}{r\rho_0}\frac{d}{dr}\left(r\rho_0 v\right) + 
          \left[\frac{i}{\sigma\rho_0}\frac{d}{dr}\left(\frac{\rho_0
	  \kappa^2}{2\Omega}\right)- \frac{im}{r}\right] u = 
          \frac{\rho_1}{\rho_0} \frac{mg_{\rm eff}}{r\sigma} \;,
     \label{dzetav}
\end{eqnarray}
where the epicyclic frequency $\kappa$ and the related vorticity
frequency $\zeta$ are defined by
\begin{eqnarray}
     \frac{\kappa^2}{2\Omega} \equiv \frac{1}{r}\frac{d}{dr}
	(r^2\Omega) \equiv 2\zeta \;.
	\label{kappa}
\end{eqnarray}

Finally, from equations~(\ref{pdivv}), (\ref{pincom}), and
(\ref{dzetav}), a wave equation can be derived for the
quantity $W\equiv ru$:
\begin{eqnarray}
     \frac{1}{r\rho_0}\frac{d}{dr}\left(r\rho_0 \frac{dW}{dr}\right) - 
	     \frac{m^2}{r^2}\left[\frac{\Omega_{\rm eff}^2}{\sigma^2}\, 
          \frac{d\ln\rho_0}{d\ln r} - \frac{2\zeta}{m\sigma}\,\frac{d\ln
		\left(\rho_0\zeta\right)
	     }{d\ln r} + 1\right] W = 0 \;.
     \label{waveeq}
\end{eqnarray}
This is the fundamental equation for the model considered in this
paper.  The mode frequency $\omega$ is an eigenvalue of this second
order differential equation and is determined by solving the equation
and applying the boundary conditions.  Any eigenvalue $\omega$ with a
positive imaginary part represents an unstable mode.  We focus on such
modes in the rest of the paper.  It is possible to derive a necessary
condition for the existence of instability, as shown in
Appendix~\ref{appa}, but we do not use that condition in the main
paper.

Once the eigenvalue $\omega$ and the eigenfunction $W\equiv ru$ are
obtained, we may solve for the other perturbed quantities by going
back to the linear perturbation equations.  Thus, equation
(\ref{pdivv}) gives the perturbed azimuthal velocity $v(r)$, and
equation (\ref{pincom}) gives the perturbed density $\rho_1(r)$.  To
calculate the perturbed total pressure $p_{t1}(r)$, we use the
azimuthal component of the first order perturbation of the momentum
equation (\ref{mom_eq}),
\begin{eqnarray}
     2\zeta u -i\sigma v = -\frac{im}{r}\,\frac{p_{t1}}{\rho_0} \;.
	\label{p_pji}
\end{eqnarray}
The perturbed magnetic field $B_1$ can be calculated from the first 
order perturbation of the induction equation~(\ref{free}),
\begin{eqnarray}
     B_1 = -\frac{i}{\sigma}\frac{dB_0}{dr}\,u \;,
	\label{B1}
\end{eqnarray}
where equation~(\ref{pdivv}) has been substituted.  Finally, the
perturbed gas pressure $p_1$ can be calculated from
equations~(\ref{p_pji}) and (\ref{B1}), by making use of $p_1 = p_{t1}
- B_0 B_1/4\pi$.

\subsection{Jump Conditions Across the Boundary}
\label{sec3.2}

Since the equilibrium model has a discontinuity at $r = r_m$,
where the accretion disk meets the magnetosphere, the solution must
satisfy certain junction conditions at this boundary.  Let us define
$r_{m+}\equiv r_m + 0^+$ and $r_{m-}\equiv r_m
+ 0^-$ as the radius $r_m$ on the two sides of the boundary.
Let us also define $\Delta_m (f)$ to denote the jump in the
value of any quantity $f$ across the boundary, i.e.
\begin{eqnarray}
     \Delta_m (f) \equiv f\left(r_{m+}\right) - f\left(r_{m-}\right) \;.
\end{eqnarray}

Consider first the Lagrangian displacement in the radial direction:
$\xi = i u/\sigma$.  Clearly, the displacement has to be the same on
the two sides of the boundary.  This gives the first jump condition:
\begin{eqnarray}
     \Delta_m \left(\frac{W}{\sigma}\right) = 0 \;.
	\label{junc1}
\end{eqnarray}
Since $\sigma$ is in general not continuous at $r = r_m$ (because
$\Omega$ need not be continuous), note that it is the quantity
$W/\sigma$ that is continuous at the boundary, while $W$ itself may or
may not be continuous, depending on the behavior of $\Omega$.

Consider next the total Lagrangian pressure perturbation at the
displaced position of the fluid:
\begin{eqnarray}
	p_{t1,{\rm L}} = p_{t1}+\xi\frac{dp_{t0}}{dr}
	     = p_{t1} - \xi\rho_0 g_{\rm eff} \;,
	\label{pt1L}
\end{eqnarray}
where we have used equation (\ref{equi_r}) in the last step.  We may
rewrite $p_{t1}$ in terms of $u$ and $v$ using equation (\ref{p_pji})
and then replace these in terms of $W$ and $dW/dr$.  Carrying out
these steps, and requiring the Lagrangian pressure perturbation to be
equal on the two sides of the boundary, we obtain the second jump
condition:
\begin{eqnarray}
     \Delta_m\left[\sigma\rho_0\, \frac{dW}{dr} -\frac{m}{r} 
	     \left(\frac{m\Omega_{\rm eff}^2}{\sigma} -2\zeta\right)\rho_0 
		W\right] = 0 \;.
	\label{junc2}
\end{eqnarray}

The two jump conditions~(\ref{junc1}) and (\ref{junc2}) form a
complete set of relations that must be satisfied at $r = r_m$.

\section{Models with Continuous $\Omega$ at the Interface}
\label{sec4}

We first consider a model in which the angular velocity is continuous
across $r_m$, and take the profile of $\Omega$ to be given by (see
Fig.~\ref{fig2})
\begin{eqnarray}
	\Omega(r) = \Omega_m\left({r\over r_m}\right)^{-q} \;.
	\label{Omega}
\end{eqnarray}
The corresponding vorticity frequency is
\begin{eqnarray}
	\zeta(r) = \left(1-\frac{q}{2}\right)\Omega(r) \;.
	\label{zetaq}
\end{eqnarray}
For completeness, we consider values of the index $q$ over the range
from 0 (constant angular velocity) to 2 (constant specific angular
momentum).  However, in real accretion disks, the relevant range is
likely to be from 3/2 (Keplerian) to 2.  We assume that the
gravitational potential $\Phi \propto r^{-1}$.

A model with continuous $\Omega$ is not likely to be relevant for
accreting neutron stars, since we expect in general a discontinuity in
$\Omega$ at the interface between the disk and the magnetosphere.
However, in the case of accretion onto a magnetosphere surrounding a
black hole, since the field is not anchored to the black hole, we
expect the magnetospheric fluid to corotate with the surrounding disk.
Thus, the discussion in this section is most relevant for the black
hole case.  It is possible that the value of $q$ may change across the
boundary for a real black hole flow.  We do not consider this
possibility here, though the equations we write are general enough to
handle it.\footnote{Neglecting a possible change in $q$ across the
boundary is reasonable especially when the following fact is
considered: when the disk has a sharp contrast in mass density across
the boundary (i.e., $\rho_+ \gg \rho_-$), the solutions do not depend
on the details of the disk inside $r = r_m$.}

As described in \S\ref{sec2.1}, we assume that there is a density jump
at $r_m$.  Thus, we model the density profile as
\begin{eqnarray}
     \rho_0 = \left\{\begin{array}{ll}
             \rho_- \;, & \mbox{if $r < r_m$} \\
             \rho_+ (r/r_m)^{-\gamma} \;, & \mbox{if $r > r_m$} 
        \end{array}\right. \;,
        \label{density2}
\end{eqnarray}
where $\rho_+$, $\rho_-$, and $\gamma$ are constants. When $\gamma>0$,
the mass density decays with increasing radius for $r > r_m$. The jump
in mass density at $r_m$ is described by the following dimensionless
parameter,
\begin{eqnarray}
     \mu\equiv \frac{\rho_+-\rho_-}{\rho_++\rho_-} \;.
	\label{mu}
\end{eqnarray}
Note that $\mu$ can in general take any value between $-1$ and $+1$.
However, we only consider values of $\mu>0$, i.e., $\rho_+ > \rho_-$.

Because $\Omega$ is continuous at $r_m$, so is $\sigma$.
Therefore, the two jump conditions written in \S\ref{sec3.2} become
simplified to
\begin{eqnarray}
	\Delta_m(W) = 0 \;,
	\label{jump1}
\end{eqnarray}
\begin{eqnarray}
     \Delta_m\left(\rho_0\frac{dW}{dr}\right) = \left[
          \frac{m^2\Omega_{\rm eff}^2}{\sigma^2} -\frac{2m\zeta}
          {\sigma}\right]_{r = r_m} \frac{W_m}{r_m}\,
          \Delta_m(\rho_0) \;. 
     \label{jump2}
\end{eqnarray}
The former equation states that $W$ is continuous at $r_m$.  We write
the value of $W$ at this radius as $W_m$.  In the latter equation,
$\rho_0=\rho_+$ for $r=r_{m+}$ and $\rho_-$ for $r=r_{m-}$.

\subsection{Analytical Results}
\label{sec4.1}

We start first with the simplest case: (1) We assume that $\gamma=0$,
i.e., the mass density is constant on the two sides of the boundary,
with a jump at the junction described by the parameter $\mu$. (2) We
assume that $q$ is equal to either 0 or 2.

The first assumption implies that the term involving $d\ln\rho_0/d\ln
r$ in equation~(\ref{waveeq}) vanishes, while the second assumption
means that the term involving $d\ln(\rho_0\zeta)/d\ln r$ vanishes,
either because $\zeta$ is constant (for $q=0$) or it is equal to 0
(for $q=2$).  Therefore, away from the boundary $r = r_m$, the wave
equation~(\ref{waveeq}) takes the form
\begin{eqnarray}
     \frac{1}{r}\frac{d}{dr}\left(r\frac{dW}{dr}\right) -\frac{m^2}
          {r^2} = 0 \;.
     \label{wave_ex1}
\end{eqnarray}
This differential equation has two simple solutions: $W \propto
r^{+m}$, $r^{-m}$.  We require the perturbations to decay away from
the boundary, both as $r\rightarrow\infty$ and as $r \rightarrow 0$.
Therefore, for $m\geq 1$, the solution must be of the form
\begin{eqnarray}
     W = \left\{\begin{array}{ll}
	     A_- r^m \;, & \mbox{if $r < r_m$}  \\
		A_+ r^{-m} \;, & \mbox{if $r > r_m$}  
	\end{array}\right. \;,
\end{eqnarray}
where $A_+$ and $A_-$ are constants. 

We now apply the two jump conditions (\ref{jump1}) and (\ref{jump2}).
Since $\Delta_m (W) =0$, we must have
\begin{eqnarray}
     A_+ = A_- r_m^{2m} \;. \label{ba}
\end{eqnarray}
Thus,
\begin{eqnarray}
     \Delta_m\left(\rho_0\frac{dW}{dr}\right) =
	  -\left(\rho_+ +\rho_-\right) m A_- r_m^{m-1} \;.
\end{eqnarray}
Applying this to the second jump condition~(\ref{jump2}), we obtain a
solution for the eigenvalue $\omega = \sigma + m\Omega$,
\begin{eqnarray}
     \omega = \left[m \Omega +\mu\zeta \pm \sqrt{-\mu m \Omega_{\rm eff}^2 
	     + \mu^2\zeta^2}\right]_{r = r_m} \;,
		\qquad q = 0,~2 \;,
	\label{omega02}
\end{eqnarray}
where $\mu$ is the density contrast parameter defined in
equation~(\ref{mu}).

We are interested in unstable modes, i.e., modes with complex
$\omega$, where the growth rate of the instability is given by
$\omega_{\rm I}$.  From equation~(\ref{omega02}), we see that unstable
modes exist whenever $-\mu m \Omega_{\rm eff}^2 +\mu^2 \zeta^2 <0$,
i.e. if
\begin{eqnarray}
      0 < \mu < \mu_1 \;, \qquad \mu_1 \equiv \left.\frac{m \Omega_{\rm
     eff}^2}{\zeta^2}\right|_{r = r_m} > 0 \;.
\end{eqnarray}
Thus, the instability exists only if $\mu$ is positive, i.e., if the
density on the outside is greater than that on the inside.  This is
perfectly natural for the Rayleigh-Taylor instability.  Surprisingly,
if the density contrast is too large, i.e., if $\mu>\mu_1$, the
instability shuts off.  This is clearly the result of rotation, or
more specifically vorticity.  When $q=2$ and the vorticity $\zeta=0$,
then $\mu_1\to\infty$, and the instability is present for any positive
value of $\mu$.  However, for $q=0$, we have $\zeta=\Omega$ and
$\mu_1$ is finite.  In this case, if $\Omega_{\rm eff}^2$ small (i.e.,
the effective gravity is weak) and if we consider a low value of the
azimuthal wavenumber $m$, the vorticity is able to eliminate the
instability.

The real part of $\omega$ gives the observed oscillation frequency of
the mode.  When $q=2$, this is simply equal to $m\Omega_m$.  In this
case, the mode is stationary in the frame of the fluid at the
boundary, and what one observes is simply the Keplerian frequency
$\Omega_m$ of the gas at the inner edge of the disk.  However, when
$q=0$, the observed mode frequency is not equal to $m\Omega_m$, but is
equal to $(m+\mu)\Omega_m$, which differs from the orbital frequency.
In this case, because of vorticity, we have a traveling mode in the
fluid frame.

\subsection{Numerical Results}
\label{sec4.2}

For a more general disk, with $\gamma\ne0$ and/or $q\ne 0,2$, we have
to solve the wave equation and the boundary conditions numerically.
Although we derived in \S\ref{sec3.1} a second-order differential
equation as our basic wave equation (eq.~[\ref{waveeq}]), for
numerical purposes a pair of equivalent first-order differential
equations is more convenient. Defining $V = r v$, we write
\begin{eqnarray}
     \frac{d}{dr}\left\{\begin{array}{c}
	       W\\[7pt]V\end{array}\right\} = \left\{\begin{array}{cc}
		       0 & ~~~-\frac{im}{r} \\[7pt]
			  \frac{im}{r}\left[1-\frac{2r}{m \sigma}\frac{d
			  \zeta}{dr}+\left(\frac{\Omega_{\rm eff}^2}{\sigma^2} 
			  -\frac{2\zeta}{m\sigma}\right)\frac{r}{\rho_0}
			  \frac{d\rho_0}{dr}\right]& 
			  ~~~-\frac{1}{\rho_0}\frac{d\rho_0}{dr} \end{array}\right\}
		       \left\{\begin{array}{c}
	            W\\[7pt]V\end{array}\right\} \;.
	\label{cdiff}
\end{eqnarray}
The first jump condition at $r = r_m$ is that $W$ must be
continuous (eq.~[\ref{jump1}]).  From equations~(\ref{jump2}) and
(\ref{pdivv}), we have
\begin{eqnarray}
     \Delta_m(\rho_0 V) = i \left[\frac{m\Omega_{\rm eff}^2}{\sigma^2}-
	     \frac{2\zeta}{\sigma}\right]_{r = r_m} 
	     W_m\,\Delta_m\rho_0 \;,
	\label{cond2}
\end{eqnarray}
where $W_m=W(r_m)$.  Substituting equation~(\ref{mu}) into this, we
obtain the second jump condition
\begin{eqnarray}
     (1+\mu) V_+ - (1-\mu) V_- = 2 i \left[\frac{m\Omega_{\rm eff}^2}
          {\sigma^2} -\frac{2\zeta}{\sigma}\right]_{r = r_m} 
	     W_m\,\mu \;.
	\label{cond2a}
\end{eqnarray}

We integrate the two first-order equations~(\ref{cdiff}) from a small
radius $r = r_1 \ll r_m$ outward, and simultaneously also from a large
radius $r = r_2 \gg r_m$ inward. Where the two integrations meet at $r
= r_m$, we adjust the relative normalization factor and the eigenvalue
$\omega$ such that the two jump conditions are satisfied.  The
starting values of $W$ and $V$ at $r = r_1$ and $r_2$ are chosen to
correspond to the appropriate eigenvectors of the operator $d/dr$ in
equation (\ref{cdiff}).  At $r = r_1$, we require that the real part
of the eigenvalue should be positive, so that $d|W|^2/dr >0$ and
$d|V|^2/dr>0$, while at $r = r_2$, we require that the real part of
the eigenvalue should be negative, so that $d|W|^2/dr <0$ and
$d|V|^2/dr <0$.  The two eigenvalues of $d/dr$ (see eq.~[\ref{cdiff}])
are
\begin{eqnarray}
     k_{1,2} = \frac{1}{2}\left\{-\frac{1}{\rho_0}\frac{d\rho_0}{dr}
	     \pm\sqrt{\left(\frac{1}{\rho_0}\frac{d\rho_0}{dr}\right)^2 +
		\frac{4m^2}{r^2}\left[1-\frac{2r}{m \sigma}\frac{d
		\zeta}{dr}+\left(\frac{\Omega_{\rm eff}^2}{\sigma^2} 
		-\frac{2\zeta}{m\sigma}\right)\frac{r}{\rho_0}
		\frac{d\rho_0}{dr}\right]}\right\} \;.
	\label{k1k2}
\end{eqnarray}
The corresponding two eigenvectors (up to a normalization factor) are,
respectively,
\begin{eqnarray}
     \left\{\begin{array}{c}
	       W\\[7pt]V\end{array}\right\}_1 = \left\{\begin{array}{c}
	            1\\[7pt] \frac{ir}{m} k_1 \end{array}\right\}\;, \qquad
	\left\{\begin{array}{c}
		  W\\[7pt]V\end{array}\right\}_2 = \left\{\begin{array}{c}
	            1\\[7pt] \frac{ir}{m} k_2 \end{array}\right\} \;.
	\label{v1v2}
\end{eqnarray}
It can be checked that, at both $r= r_1 \ll r_m$ and $r= r_2 \gg r_m$, 
$k_1$ is positive and $k_2$ is negative.  Therefore, the eigenvector
corresponding to $k_1$ is a physical solution at $r = r_1$, and the
eigenvector corresponding to $k_2$ is a physical solution at $r =
r_2$.  Thus, in our numerical calculations, we choose $W = r_1^m$ and
$V = i r_1^{m+1} k_1(r_1)/m$ at $r = r_1$, and $W = r_2^{-m}$ and $V =
i r_2^{-m+1} k_2(r_2)/m$ at $r = r_2$, as the starting values of $W$
and $V$.

One comment is in order regarding the integration.  The point
$\sigma=0$ is a singularity of equation (\ref{cdiff}); it represents
the corotation singularity.  Therefore, while integrating the equation
numerically, it is important to make sure that the integration contour
goes above the singularity \citep{nar87}.  When $\omega$ (thus
$\sigma$) is complex, this is easily arranged by just integrating
along the real $r$-axis since the singularity is below the real axis
for positive $\omega_{\rm I}$ (which corresponds to an unstable mode,
recall our convention that the perturbations are $\propto
\exp[-i\omega t]$).  However, when $\omega$ is real, the corotation
singularity is on the real $r$-axis.  In this case, the integration
contour must go above the singularity in the complex $r$-plane.

We present some numerical results in
Figs.~\ref{fig3}--\ref{fig5}. In Figure~\ref{fig3}, we show the
frequency $\omega = \omega_{\rm R} + i \omega_{\rm I}$, in units of
$\Omega_m = \Omega (r_m)$, corresponding to a model with $\mu = 1$,
$\Omega_{{\rm eff},m}^2 = 0.6 \Omega_m^2$, and $\gamma=0$ (i.e.,
constant mass density away from $r = r_m$).  The solid lines show
$\omega_{\rm I}$ and give the growth rate of unstable modes. The
dashed lines represent $\omega_{\rm R}$, and correspond to the
observed oscillation frequency of the mode.  The results are perfectly
reasonable --- mode frequencies vary smoothly as a function of $q$
between the two limits $q=0$ and $q=2$ (where we have analytical
results, \S\ref{sec4.1}).  For $m=1$ and small values of $q$, we see
that the instability shuts off and we have instead two neutral modes
with real $\omega$.  This is not unexpected.  The analytical results
in \S\ref{sec4.1} indicate that for $q=0$ and this set of parameters,
we have $\mu>\mu_1$, which indicates that there should be no
instability.

For the parameters used in Fig.~\ref{fig3} and for $q$ in the range
1.5--2 of interest for accretion disks, we find that all azimuthal
wave numbers $m$ are unstable.  However, as equation (37) shows,
whether or not a given mode is stable or unstable depends on whether
$\mu$ is greater than or less than $\mu_1$ and this depends on the
magnitude of $\Omega_{\rm eff}^2 /\zeta^2$.  (Although the result
given in eq.~[37] is true only for $q=0,~2$, the qualitative result is
valid also for other values of $q$.)  In Figure~\ref{fig3a} we show
the numerically determined critical values of $\Omega_{{\rm eff},
m}^2$ that separate stable and unstable modes in the $q-\Omega_{{\rm
eff},m}^2$ plane (we restrict our attention to $q$ in the range
$1.5$--$2$ which is of most interest for QPOs).  The results
correspond to $\mu =1$, $\gamma =0$. We see that, for $q<2$, there
always exists a range of $\Omega_{{\rm eff},m}^2$ over which modes
with a given $m$ are stable.  For example, when $q = 1.5$ and
$0.024<\Omega_{{\rm eff},m}^2/\Omega_m^2<0.04$, modes with $m = 1$ are
stable and modes with $m\geq2$ are unstable, while for
$0.018<\Omega_{{\rm eff},m}^2/\Omega_m^2<0.024$, modes with $m \leq 2$
are stable and modes with $m\geq3$ are unstable.

Figure~\ref{fig4} shows results for some other choices of parameters:
$\mu = 0.8$, $\Omega_{{\rm eff},m}^2 = \Omega_m^2$, and $\gamma = 0$
in panel (a); $\mu = 1$, $\Omega_{{\rm eff},m}^2 = \Omega_m^2$, and
$\gamma = 0$ in panel (b); and, $\mu = 1$, $\Omega_{{\rm eff},m}^2 =
\Omega_m^2$, and $\gamma = 1$ in panel (c). Comparing panel (b) with
Figure~\ref{fig3}, we see that the zone of stable modes at small $q$
and $m=1$ seen in the latter, disappears when $\Omega_{\rm eff}^2$
increases.  This is because $\mu_1$ increases and becomes larger than
$\mu$.  All other features are very similar.  Comparing panels (a) and
(b) of Figure~\ref{fig4}, we see the effect of varying $\mu$.  For this
particular set of parameters, for $m = 1$ we see that $\omega_{\rm I}$
decreases with increasing $\mu$ when $q$ is close to 0, but increases
with increasing $\mu$ when $q$ is close to 2.  For $m>1$, however,
$\omega_{\rm I}$ always increases with increasing $\mu$. The real part
of the frequency, $\omega_{\rm R}$, increases with increasing $\mu$
for $0\leq q < 2$, but remains independent of $\mu$ for $q = 2$ (see
eq.~[\ref{omega02}]).  Finally, by comparing panels (b) and (c), we
see the effect of having a non-constant density ($\gamma\ne0$).
Generally, as $\gamma$ increases, $\omega_{\rm I}$ also increases, but
$\omega_{\rm R}$ is hardly affected.

In Figure~\ref{fig5} we show results corresponding to a Keplerian-type
disk ($q = 3/2$), as a function of $\mu$, for $\Omega_{{\rm eff},m}^2 =
0.2 \Omega_m^2$ and $\gamma=1$ (i.e., the mass density decays
with increasing radius according to $\rho\propto r^{-1}$). We see that
the imaginary part of the frequency $\omega_{\rm I}$ strongly depends
on $\mu$, with $\omega_{\rm I}\rightarrow 0$ as $\mu\rightarrow
0$. This is because the instability is driven by the density contrast
at the boundary.  As the density contrast goes to zero, the
instability must clearly vanish.  The real part of the frequency,
$\omega_{\rm R}$, however, depends only weakly on $\mu$.

Summarizing, we find that the Rayleigh-Taylor instability operates
over a wide range of parameters so long as the density on the outside
is greater than the density on the inside.  The growth rate of the
instability is typically large, of order $\sim\Omega_{\rm eff}$, which
means that the growth occurs fairly rapidly.  For $3/2\le q\le2$, the
range of interest for accretion disks, the real part of the frequency
is nearly equal to $m\Omega_m$ in almost all cases.  This is because
the mode nearly corotates with the gas at $r_m$.  For a large density
contrast $\mu\to1$ and a weak effective gravity $\Omega_{\rm eff}^2
\ll \Omega_m^2$, low-$m$ modes become stable and the instability is
limited to higher-$m$ modes.

\section{Models with Discontinuous $\Omega$ at the Interface}
\label{sec5}

We now consider a model that is more appropriate for an accretion flow
around a magnetized neutron star (Ghosh \& Lamb 1978, 1979).  Since
the magnetic field of a neutron star is frozen to the star, the low
density plasma in the magnetosphere ($r < r_m$) must corotate rigidly
with the star.  However, outside the magnetospheric radius, we expect
the accreting gas to rotate differentially as in a standard accretion
disk.  Also, the angular velocity of the disk at $r_m$ will, in
general, not match the angular velocity of the star.  Based on these
considerations, we assume the following model for the angular velocity
(Fig.~\ref{fig6}):
\begin{eqnarray}
     \Omega = \left\{\begin{array}{ll}
             \Omega_- \;, & \mbox{if $r < r_m$} \\
             \Omega_+ (r/r_m)^{-q} \;, & \mbox{if $r > r_m$} 
        \end{array}\right. \;,
\end{eqnarray}
where $\Omega_+$ and $\Omega_-$ are constants. In general $\Omega_-
\neq \Omega_+$.  For the density profile, we assume the same model as
in equation~(\ref{density2}).
	
The jump conditions are given by equations~(\ref{junc1}) and
(\ref{junc2}).

\subsection{Analytical Results}
\label{sec5.1}

As in \S\ref{sec4.1}, we consider the special case when $\gamma=0$ and
$q=0$ or 2.  We may then repeat the same steps described in
\S\ref{sec4.1}, except that the jump conditions are now different.  We
obtain the following results for the eigenvalue,
\begin{eqnarray}
     \omega = \frac{1}{2}\left[(1+\mu)\left(m\Omega_+ + \zeta_+\right) + 
	     (1-\mu)\left(m\Omega_- - \zeta_-\right) \pm \sqrt{-\Theta}
	     \right] \;,
	\label{omega_e}
\end{eqnarray}
where the ``$+$'' and the ``$-$''in the subscripts denote evaluations
at $r = r_{m+}$ and at $r =r_{m-}$, respectively, and
\begin{eqnarray}
     \Theta &\equiv& 2\left[(1+\mu)\left(m \Omega_{{\rm eff},+}^2 - 
		\zeta_+^2\right) -(1-\mu)\left(m \Omega_{{\rm eff},-}^2  
		+\zeta_-^2\right)\right] \nonumber\\
		&&+\left(1-\mu^2\right)\left[m\left(\Omega_+ -\Omega_-\right)+
		\zeta_+ + \zeta_-\right]^2 \;.
	\label{theta}
\end{eqnarray}
When $\Omega_+ =\Omega_-$ (which means $\zeta_+ =\zeta_-$,
$\Omega_{{\rm eff},+} = \Omega_{{\rm eff},-}$),
equation~(\ref{omega_e}) simplifies to equation~(\ref{omega02}).

When $\Theta$ is positive, there is an instability.  A study of the
expression for $\Theta$ indicates that there are two distinct terms,
written in separate lines in equation (\ref{theta}).  Therefore, there
are two distinct mechanisms of instability.  The first term in
(\ref{theta}) is the one we have already seen in \S\ref{sec4}, due to
the Rayleigh-Taylor instability.  If $\Omega$ and its derivative are
continuous at $r_m$, this is the only term present, and it gives an
instability only if $\rho_+>\rho_-$, i.e., $\mu>1$.  However, for the
problem at hand, $\Omega$ is discontinuous at $r_m$, and so the second
term can also be important.  Ignoring the vorticity terms
$\zeta_{\pm}$ for simplicity, this term is directly proportional to
$(\Omega_+-\Omega_-)^2$, i.e., the square of the angular velocity jump
across the boundary.  This term is nothing but the classical
Kelvin-Helmholtz instability associated with a velocity discontinuity.

The observed mode frequency is given by the real part of $\omega$ in
equation~(\ref{omega_e}).  If $\mu=1$, i.e., we have the maximum
density contrast across the boundary, then the frequency is just equal
to the appropriate frequency $m\Omega_++\zeta_+$ of the outer
accretion disk at $r_{m+}$.  That is, the rotation rate of the neutron
star $\Omega_-$ is irrelevant.  This is reasonable since there is no
mass associated with the flow inside $r_m$, and so the frequency
$\Omega_-$ of this region of the flow should have no influence on the
mode.  However, when $\mu\ne1$, the mode frequency is a linear
combination of $m\Omega_++\zeta_+$ and $m\Omega_-+\zeta_-$, with
weights given by the two densities.

All of these results are very reminiscent of the classical results for
the Rayleigh-Taylor and Kelvin-Helmholtz instability \citep{dra81}.
We discuss this connection in Appendix~B.

\subsection{Numerical Results}
\label{sec5.2}

For the general case with $\gamma \neq 0$ and/or $q \neq 0,2$, we
numerically solve equations~(\ref{cdiff}), (\ref{junc1}), and
(\ref{junc3}), following the numerical approach described in
\S\ref{sec4.1}.  We write the jump condition~(\ref{junc2}) as
\begin{eqnarray}
     \Delta_m\left[\sigma\rho_0 V -i \left(\frac{m\Omega_{\rm eff}^2}
	     {\sigma} -2\zeta\right)\rho_0 W\right] = 0 \;.
	\label{junc3}
\end{eqnarray}

In Figure~\ref{fig7} we show some numerical results for a disk with
$\gamma=0$, $\mu=0.4$ and the frequency of effective gravity
$\Omega_{{\rm eff},+}^2= \Omega_+^2$. In panel (a), the disk angular
velocity satisfies $\Omega_- = 2 \Omega_+$, i.e., the neutron star
rotates twice as fast as the disk at $r=r_m$. In panel (b), the disk
angular velocity satisfies $\Omega_- = \Omega_+$, i.e., the neutron
star rotates at the same rate as the disk and the angular velocity is
continuous (but not smooth) at $r = r_m$. In panel (c), the disk
angular velocity satisfies $\Omega_- = 0.4 \Omega_+$, i.e., the
neutron star rotates more slowly than the disk.

Comparing the three panels, we see that the real part of the frequency
($\omega_{\rm R}$) monotonically decreases as $\Omega_-/\Omega_+$
decreases.  This is as expected.  Based on the discussion in
\S\ref{sec5.1}, the mode frequency is a linear combination of the
outer and inner frequencies at the boundary.  As the latter decreases,
the mode frequency must also decrease.  The growth rate of the mode is
more interesting; it is minimum when $\Omega_-/\Omega_+ =1$, and
increases both for $\Omega_-/\Omega_+>1$ and $\Omega_-/\Omega_+<1$.
This behavior is because of the simultaneous presence of the
Rayleigh-Taylor and Kelvin-Helmholtz instabilities.  When
$\Omega_-/\Omega_+=1$, there is no velocity discontinuity at the
boundary and we have only the Rayleigh-Taylor instability, with a
certain growth rate.  However, both for $\Omega_-/\Omega_+<1$ and
$\Omega_-/\Omega_+>1$, there is an additional contribution to the
growth rate from the Kelvin-Helmholtz instability, and so the net
growth rate is enhanced.

Note that, for the case shown in panel (a), we have $\Omega_{{\rm
eff},-}^2 = -0.5 \Omega_-^2$, indicating that the disk is
super-Keplerian right inside $r = r_m$. But there are still unstable
modes in this case, since the disk is sub-Keplerian for $r>r_m$. For
the cases shown in panels (b) and (c), the disk is sub-Keplerian for
both $r < r_{m}$ and $r > r_{m}$.

\section{Summary and Discussion}
\label{sec6}

The main message of this paper is that there are instabilities
associated with the magnetospheric radius where an accretion disk
meets the magnetosphere of the central mass and that these
instabilities may be relevant for understanding QPOs in binary
systems.  There are two natural instabilities at the
disk-magnetosphere interface: (1) Rayleigh-Taylor (or interchange)
instability associated with a density jump, and (2) Kelvin-Helmholtz
instability associated with an angular velocity jump.  These
instabilities, which are well-known for classical uniform Cartesian
flows (Appendix B), survive with relatively little modification for a
rotating, shearing flow with a density gradient (\S\S4,5).  The
unstable modes are expected to grow to become nonaxisymmetric
perturbations with large amplitude (e.g., the streams in the
simulations of Igumenshchev et al. 2003) and to give strong
quasi-periodic variations in the observed intensity.

If, as we propose, these instabilities are responsible for the
observed QPOs in X-ray binaries, the same model might work both for
neutron stars and black holes.  The existence of a disk-magnetosphere
interface around accreting magnetized neutron stars is well-known.
The popular beat frequency model for QPOs invokes blobs in the
accretion disk orbiting at the Keplerian frequency at the
magnetospheric radius \citep{alp85,lam85,str96} or at the sonic radius
\citep{mil98,lam01}. While the model does not explain the origin of
the blobs, it is reasonable to assume that the blobs are in some cases
at least created by the interchange and Kelvin-Helmholtz instabilities
studied in this paper.  A fact not widely appreciated is that the
magnetospheric model might also apply to accreting black holes and
unmagnetized neutron stars.  As Bisnovatyi-Kogan \& Ruzmaikin (1976)
showed, it is possible for the region close to a black hole to become
magnetically dominant (Livio et al. 2003; Narayan et al. 2003).  An
accretion disk would then be disrupted at a magnetospheric radius,
just as in the neutron star case, and the interface would be unstable
and produce QPOs.  The main difference between the two cases is that
the magnetic field is not anchored to the black hole, and so the
magnetosphere does not rotate rigidly with the star as in the neutron
star case.  This difference between the black hole and neutron star
problems leads to some differences in the results for the two cases,
as discussed in \S\S4,5.

Our analysis shows that nearly all azimuthal wavenumbers $m$ are
unstable under reasonable conditions.  Although modes with higher
values of $m$ grow more rapidly, we expect that these modes will
saturate at relatively small amplitudes.  The low-$m$ modes, on the
other hand, are likely to grow to large amplitude and are therefore of
most interest for understanding QPOs.  For $q$ in the range 3/2 to 2
(Keplerian to constant angular momentum), and reasonable assumptions
about the density profile, we have shown that the observed mode
frequency tends to be of order $m \Omega_m$.  Thus, in the simplest
version of the model, we expect the observed QPO frequencies to be in
the ratio 1:2:3, etc.  However, as we showed in \S\ref{sec4} (see
Fig.~4), it may often be the case that the $m=1$ mode is stable and
that only modes with $m\geq2$ are unstable.  This should happen
whenever the effective gravity in the radial direction is weak, i.e.,
when the gas pressure in the disk is low.  In this case, the QPO
frequencies should be roughly in the ratio 2:3:4, etc.  It is
interesting that a frequency ratio 2:3 is frequently seen in both
black hole and neutron star systems \citep{abr01,abr03,rem02a}.

We have emphasized that, in our opinion, QPOs should be produced by
the growing modes in the disk. Our belief is based on the fact that
the amplitude of intensity fluctuations observed in QPOs is often
quite large. Observations also show that QPOs have a finite frequency
width, which indicates that the unstable modes in the disk have a
finite life time (hence the term quasi-periodic oscillations). Apart
from the rotation period, there are two natural time scales in the
disk: the time scale associated with the effective gravity, which is
$\sim 1/\Omega_{\rm eff}$, and the viscous time scale, which is $\sim
r/v_r$, where $v_r$ is the mean radial velocity of the disk fluid.
For a standard disk, $\Omega_{\rm eff} \sim c_s/r$, where $c_s$ is the
sound speed.  Since $c_s \gg v_r$, the time scale $1/\Omega_{\rm eff}$
is generally much shorter than $r/v_r$.  Therefore, the likely
lifetime of blobs created by the instability is $1/\Omega_{\rm eff}$.
Once a mass blob is formed, it will drift toward the central object
under the action of gravity, causing a displacement in the center
frequency and a width to the QPO feature in the power spectrum. The
width is likely to be $\Delta f \sim \Omega_{\rm eff}/2\pi \sim
(h/r)(\Omega_m/2\pi)$, where $h$ is the vertical thickness of the 
disk, and the displacement speed is approximately $df/dt \sim 
(\Omega_{\rm eff}/2\pi) f$.
 
Another issue concerns how the presence of a nonaxisymmetric mode
translates to a time modulation of the observed flux.  Two
possibilities are likely.  One is that the system is viewed in a
nearly edge-on configuration so that the accreting star eclipses the
far side of the disk.  Then, as bright and faint segments of the disk
are successively eclipsed the signal at the observer will be
modulated.  The other possibility, which also requires fairly high
inclination, is that the motion of the gas is relativistic and the
observed signal is dominated by the blue-shifted segment of the disk.
Once again, as the nonaxisymmetric pattern rotates, the signal will
oscillate.  Both mechanisms require that the bright and faint patches
on the disk should have large areas, since otherwise the fractional
modulation of the observed X-ray flux will be small.

The azimuthal extent of a bright patch in a nonaxisymmetric mode with
wavenumber $m$ is $\sim\pi/m$.  The radial extent also has an
$m$-dependence, since away from the magnetospheric radius the 
perturbation solutions decay with radius as $\sim r^{\pm m}$ (\S4.1).  
For both reasons, low-$m$ modes have patches with the largest area and 
hence are most promising.  The area occupied by a bright patch, 
defined to correspond to an annular region in the disk bounded by a 
radius where the perturbation amplitude is half the peak, is estimated 
to be $S \sim (\pi r_m^2/2m) (4^{1/m}-4^{-1/m})$.  For $m
=1, 2, 3$ we have $S \sim 1.875\pi r_m^2, 0.375 \pi r_m^2, 0.16 \pi r_m^2$,
respectively. For large $m$, $S$ approaches zero according to $S\sim
2\pi r_m^2 (\ln 2/m^2)$.  The scaling clearly shows that low-$m$ modes
dominate by a large factor.  In addition, as we argued earlier,
low-$m$ modes are likely to saturate with substantially larger
amplitudes than high-$m$ modes.  This is yet another reason why only
the lowest order few modes are expected to cause a discernible signal
in the observations.

Apart from demonstrating that unstable modes exist and that mode
frequencies in the ratio 2:3 are possible, the model does not really
explain any of the many puzzling features seen in the observations
(see the summary in \S\ref{sec1}).  For instance, the model does not
explain why the frequency difference between twin kHz QPOs in neutron
star systems is often roughly of order the neutron star spin frequency
(van der Klis 2000) or sometimes half the spin frequency (Wijnands et
al. 2003).  Even though the version of the model described in \S5
appears to have the necessary ingredients for the beat frequency model
to operate, namely gas orbiting at Keplerian frequency around a
magnetosphere that rotates at the stellar frequency, nevertheless our
analysis does not reveal any beat phenomenon.  Clearly, additional
physics is needed beyond what we have considered here.

A Rayleigh-Taylor-like process has been studied extensively by
Titarchuk and collaborators in a sub-Keplerian transition region of a
disk around a black hole or a weakly magnetized neutron star
\citep{tit98,osh99,tit99,tit02,tit03}.  These authors focus only on
stable modes and suggest that their model can explain the observed
correlation between the twin kilohertz frequencies and the horizontal
branch QPO frequency \citep{osh99,tit03}.  In their model, the
dynamical effect of the magnetic field is always assumed to be
unimportant, so the fluid is described purely within hydrodynamics.
The model assumes the existence of a thin sub-Keplerian transition
region in the vicinity of the compact central object where the
accreting matter adjusts itself either to the surface of a rotating
neutron star or to the innermost boundary of the accretion disk
\citep{tit98}. But the origin of the transition layer is not
explained, especially considering that the magnetic field is assumed
to be weak.

The model described in the present paper (see \S\ref{sec2.1}) differs
from Titarchuk's model in two respects.  First, we assume that the
magnetic field is dynamically important inside the inner edge of the
disk.  Second, we focus on genuinely unstable modes rather than on
stable modes, since we assume that only unstable modes can grow to a
large enough amplitude to produce the observed intensity
fluctuations. Because of the assumption of a strong magnetic field, a
narrow transition region develops naturally at the disk-magnetosphere
interface of the model, and unstable oscillations are triggered at
this interface.

\citet{kai92} used two-dimensional shearing box simulations to study
the magnetic interchange instability in an accretion disk with a
vertical magnetic field.  When the magnetic field is strong and its
strength decreases with increasing radius, they find that an
instability develops spontaneously.  The model we consider is an
extreme version of the Kaisig et al. model in which the magnetic field
decreases discontinuously at the magnetospheric radius. Our analytical
results are consistent with \citet{kai92}'s numerical result that, in
the linear regime, the growth rate of perturbations depends on the
azimuthal wave number $k=m/r$ of the initial perturbations as
$\omega_{\rm I} \sim k^{1/2}$.  In both studies, the dependence of
quantities on $z$ is neglected. When this dependence is included,
\citet{lov94} showed that the twisting of field lines acts to drive
winds or jets from the disk surfaces, which increases the disk
accretion speed and so amplifies the magnetic field and leads to
runaway or implosive accretion and explosive wind or jet
formation. Similar results have also been obtained by Lubow,
Papaloizou \& Pringle (1994a,b). [However, Lubow et al. 1994b also
claimed the existence of stable solutions with no accretion and no
wind.]

\citet{lub95} and \citet{spr95} found that disk shear tends to
stabilize the disk-magnetosphere interface. This seems to be in
conflict with our result that the growth rate of the Rayleigh-Taylor
instability increases with increasing $q$. However, we stress that in
their model the magnetic field has both vertical and radial
components.  Indeed, they modeled the disk as a sheet of zero
thickness, so that the radial component of the magnetic field has a
jump as one goes from below to above the disk. The pressure of gas and
magnetic fields is negligible in the disk, but the magnetic curvature
force appears in the equation of motion. This is very different from
our model, where the disk has a nonzero (indeed infinite) thickness,
the magnetic field has only a vertical component, the curvature force
of the magnetic field is zero, and the magnetic pressure and gas
pressure both play a dynamical role.  Our model is perhaps more
suitable for describing the central layer of the disk, while their
model may be more applicable to the surface layers.

\cite{cha01} has considered the Rayleigh-Taylor instability of a
strong vertical magnetic field confined by a disk threaded with a
horizontal magnetic field.  The aim of his study was to understand the
equilibrium of magnetic flux tubes observed in the central regions of
the Galaxy.  Although the model has some points of similarity with the
present study, there are also large differences.  The disk in
Chandran's model is assumed to have a uniform rotation and the
gravitational potential is assumed to correspond to a constant
background mass density (so that the gravitational acceleration
increases linearly with radius).  In contrast, we assume that the disk
is differentially rotating and that the gravitational potential
corresponds to that of a compact mass at the center.  However,
Chandran considers a compressible gas whereas we simplify our problem
by taking the gas to be incompressible.

\citet{cha01} has derived a formula (his eqs.~[91] and [92]) for the
oscillation frequency when there is no magnetic field and the density
contrast parameter $\mu =\pm 1$.  His results are consistent with our
analytical results for a disk with constant angular velocity (see our
eq.~[\ref{omega02}]). In particular, the results confirm that the disk
vorticity has the effect of stabilizing the modes.

\acknowledgements

The authors thank the anonymous referee for several useful comments.  
LXL's research was supported by NASA through Chandra Postdoctoral 
Fellowship grant number PF1-20018 awarded by the Chandra X-ray Center, 
which is operated by the Smithsonian Astrophysical Observatory for 
NASA under contract NAS8-39073.  RN was supported in part by NASA grant
NAG5-10780 and NSF grant AST 0307433.

\begin{appendix}
\section{A Necessary Condition for Disk Instability}
\label{appa}

A necessary condition for instability can be derived from the wave
equation~(\ref{waveeq}), following the approach pioneered by
\citet{how61} (see also Li et al. 2003, Appendix B).

Defining $W = \varphi\sqrt{\sigma}$, equation~(\ref{waveeq}) can be
recast as
\begin{eqnarray}
     \frac{d}{dr}\left(r\rho_0\sigma\frac{d\varphi}{dr}\right)
	     +\frac{m^2}{r^2} \left\{-\left[\frac{d\ln\rho_0}{d\ln r}\,
		\Omega_{\rm eff}^2+ \frac{1}{4}\left(r\frac{d\Omega}{dr}
		\right)^2\right]\frac{1}{\sigma} \right. &&\nonumber\\
		\left.-\sigma+ \frac{2\zeta}m\frac{d\ln\left(\rho_0
		\Omega_\kappa\right)}{d\ln r}- \frac{r}{2m\rho_0}
		\frac{d}{dr}\left(r\rho_0\frac{d\Omega}{dr}\right)
		\right\} r\rho_0 \varphi &=& 0 \;.
	\label{phieq1}
\end{eqnarray}
Multiplying equation~(\ref{phieq1}) by the complex conjugate variable
$\varphi^*$ and integrating over the whole flow yields
\begin{eqnarray}
     \int_{r_1}^{r_2} \left(\sigma\left|\frac{d\varphi}{dr}
	     \right|^2 +\frac{m^2}{r^2} \left\{\left[\frac{d\ln\rho_0}
		{d\ln r}\,\Omega_{\rm eff}^2+ \frac{1}{4}\left(r\frac{d\Omega}
		{dr}\right)^2\right]\frac{1}{\sigma}+\sigma\right.\right. 
		&&\nonumber\\
		\left.- \frac{2\zeta}{m}\frac{d\ln\left(\rho_0\zeta\right)}
		{d\ln r}\left.+ \frac{r}{2m\rho_0}\frac{d}{dr}\left(r\rho_0
		\frac{d\Omega}{dr}\right)\right\}\left|\varphi\right|^2
		\right)r\rho_0 dr &=& \left.r\rho_0\sigma\varphi^*
		\frac{d\varphi}{dr}\right|_{r_1}^{r_2} \;.
	\label{iphieq1}
\end{eqnarray}

Taking the boundary condition to be $W = 0$ at $r_1$ and $r_2$, we see
that the right-hand side of equation~(\ref{iphieq1}) vanishes. Then,
the imaginary part of the left-hand side of equation~(\ref{iphieq1})
must vanish\footnote{If the boundary condition is taken to be $dW/dr
=0$, then it can be shown that the right-hand side of
eq.~(\ref{iphieq1}) is real, and the result is still the same.}
\begin{eqnarray}
     \omega_{\rm I} \int_{r_1}^{r_2} \left(\left|\frac{d\varphi}{dr}
	     \right|^2 +\frac{m^2}{r^2} \left\{-\left[\frac{d\ln\rho_0}
		{d\ln r}\,\Omega_{\rm eff}^2+ \frac{1}{4}\left(r\frac{d\Omega}
		{dr}\right)^2\right]\frac{1}{\left|\sigma\right|^2}+ 
		1\right\} \left|\varphi\right|^2
		\right)r\rho_0 dr = 0 \;,
	\label{iphieq2}
\end{eqnarray}
where $\omega_{\rm I} \equiv \Im(\omega) = \Im(\sigma)$. The integral
in equation~(\ref{iphieq2}) is positive, and therefore $\omega_{\rm I}
= 0$, unless the condition,
\begin{eqnarray}
     R_{\rm eff} < \frac{1}{4} \;,
	\label{condition}
\end{eqnarray}
is satisfied somewhere in the region of integration, where
\begin{eqnarray}
     R_{\rm eff} \equiv -\frac{d\ln\rho_0}{d\ln r} \left(r\frac{d\Omega}
		{dr}\right)^{-2}\,\Omega_{\rm eff}^2 = -\left(\frac{\Omega_{\rm
		eff}}{\Omega}\right)^2\left(\frac{d\ln\Omega}{d\ln r}\right)^{-2}
		\frac{d\ln\rho_0}{d\ln r} \;.
	\label{rg}
\end{eqnarray}
Note that $\Omega_{\rm eff}^2$ can be either positive or negative.

Therefore, a necessary condition for the existence of instability
(i.e., $\omega_{\rm I}\neq 0$) is that the
inequality~(\ref{condition}) must hold somewhere in the
disk. Obviously, the converse of the inequality, i.e.  $R_{\rm eff}
\geq 1/4$, is a sufficient condition for stability.

\section{Classical Rayleigh-Taylor and Kelvin-Helmholtz Instability}
\label{appb}

Let us start from the hydrodynamic equations~(\ref{cont_eq}) and
(\ref{mom_eq}) (with ${\bf B} =0$), and use Cartesian coordinates
$(x,y,z)$.  If the fluid is incompressible, equation~(\ref{cont_eq})
becomes equations~(\ref{divv}) and (\ref{incom}). Let us assume that
the system is two-dimensional and that all quantities are independent
of $z$.  We take gravity to be in the $x$ direction: ${\bf g}=
-\nabla\Phi(x) = -g(x) {\bf e}_x$. For the unperturbed equilibrium
state we assume that $\rho = \rho_0(x)$, $p = p_0(x)$, and ${\bf v} =
V(x) {\bf e}_y$. Then, the equilibrium in the $x$-direction is given
by
\begin{eqnarray}
     g = -\frac{1}{\rho_0}\frac{d p_0}{dx} \;.
\end{eqnarray}

Consider now linear perturbations 
\begin{eqnarray}
     \left\{\begin{array}{c}
              \delta\rho \\ \delta p \\ \delta{\bf v}
              \end{array}\right\}
             = \left\{\begin{array}{c}
                 \rho_1(x) \\ p_1(x) \\ u(x)\, {\bf e}_x + v(x)\, 
				 {\bf e}_y\end{array}\right\}
                \times e^{-i\omega t + ik_y y} \;,
             \label{drho_a}
\end{eqnarray}
where $\omega$ and $k_y$ are constants.  The first order perturbations
of equations~(\ref{divv}) and (\ref{incom}) then give
\begin{eqnarray}
     \frac{du}{dx} + ik_y v &=& 0 \;, \label{de1}\\
	i\sigma - u \frac{d\rho_0}{dx} &=& 0 \;,\label{de2}
\end{eqnarray}
respectively, where
\begin{eqnarray}
     \sigma\equiv \omega - k_y V \;.
\end{eqnarray}
The first order perturbation of equation~(\ref{cmom_eq}), which is
derived from equation~(\ref{divv}) and the curl of
equation~(\ref{mom_eq}), leads to
\begin{eqnarray}
     \frac{1}{\rho_0}\frac{d}{dx}\left(\rho_0 v\right) + 
          \left[\frac{i}{\sigma\rho_0}\frac{d}{dx}\left(\rho_0\frac{dV}
		{dx}\right)- i k_y\right] u = 
          -\frac{\rho_1}{\rho_0} \frac{k_y}{\sigma} g \;.
		\label{de3}
\end{eqnarray}

>From equations~(\ref{de1})--(\ref{de3}) we can derive a wave equation
for $u$,
\begin{eqnarray}
     \frac{1}{\rho_0}\frac{d}{dx}\left(\rho_0 \frac{du}{dx}\right) - 
             k_y^2\left[\frac{g}{\sigma^2 x}\, 
          \frac{d\ln\rho_0}{d\ln x} - \frac{2\zeta}{k_y x \sigma}\,\frac{d\ln
                \left(\rho_0\zeta\right)
             }{d\ln x} + 1\right] u = 0 \;,
     \label{waveeq_a}
\end{eqnarray}
where the vorticity $\zeta$ is given by
\begin{eqnarray}
     \zeta \equiv \frac{1}{2}\frac{dV}{dx} \;.
\end{eqnarray}

As in the main paper, we can derive the following two junction
conditions at a surface of discontinuity at $x =0$:
\begin{eqnarray}
     \Delta_m \left(\frac{u}{\sigma}\right) = 0 \;,
        \label{junc1_a}
\end{eqnarray}
\begin{eqnarray}
     \Delta_m\left[\sigma\rho_0\, \frac{du}{dx} -k_y 
             \left(\frac{k_y g}{\sigma} -2\zeta\right)\rho_0 
                u\right] = 0 \;.
        \label{junc2_a}
\end{eqnarray}
where $\Delta_m(f) \equiv f(x=0^+)-f(x=0^-)$.

Following Drazin \& Reid (1981, Chapter 4, except that we switch $x$ and $z$
to be consistent with the convention in this paper), we assume for the
unperturbed equilibrium state
\begin{eqnarray}
     \rho = \left\{\begin{array}{ll} \rho_- \;, & \mbox{if $x < 0$} \\
             \rho_+ \;, & \mbox{if $x > 0$} \end{array}\right. \;, \qquad
             {\bf v} = \left\{\begin{array}{ll} V_- \,{\bf e}_y \;, &
             \mbox{if $x < 0$} \\ V_+ \,{\bf e}_y \;, & \mbox{if $x >
             0$} \end{array}\right. \;,
\end{eqnarray}
where $\rho_-$, $\rho_+$, $V_-$, and $V_+$ are constants. Then, on
each side of $x=0$, we have $\zeta =0$ and $d\rho/dx =0$, and the
solutions of the wave equation~(\ref{waveeq_a}) are $u \propto
\exp(\pm k_y x)$. Requiring that $u(x\rightarrow\pm\infty) =0$, then
from the junction conditions~(\ref{junc1_a}) and (\ref{junc2_a}) we
can solve for the eigenvalue for the frequency (assuming that $k_y>0$)
to obtain
\begin{eqnarray}
     \omega = k_y \frac{\rho_+ V_+ + \rho_- V_-}{\rho_+ + \rho_-} \pm
	     \sqrt{-\frac{\rho_+-\rho_-}{\rho_+ +\rho_-}\,k_y g -
		\frac{\rho_+\rho_-}{\rho_+ +\rho_-}\,k_y^2 \left(V_+-V_-
		\right)^2} \;.
	\label{freq_a}
\end{eqnarray}

Equation~(\ref{freq_a}) is exactly the same as equation~(4.20) in
\citet{dra81}, provided we substitute their $s$ with our $-i\omega$,
and their subscripts ``1'' and ``2'' with our subscripts ``$-$'' and
``$+$'', respectively. The first term under the square root in
equation~(\ref{freq_a}) represents the Rayleigh-Taylor instability,
while the second term represents the Kelvin-Helmholtz instability.

The analysis presented in the main paper goes beyond this classical
analysis in a few respects.  We include rotation (and correspondingly
vorticity), our flow has shear (actually differential rotation), and
we consider a density gradient.  These additional features modify some
of the results, but the basic physics remains the same, namely the
presence of the Rayleigh-Taylor and Kelvin-Helmholtz instabilities.

\end{appendix}


\clearpage
\begin{figure}
\epsscale{1}
\plotone{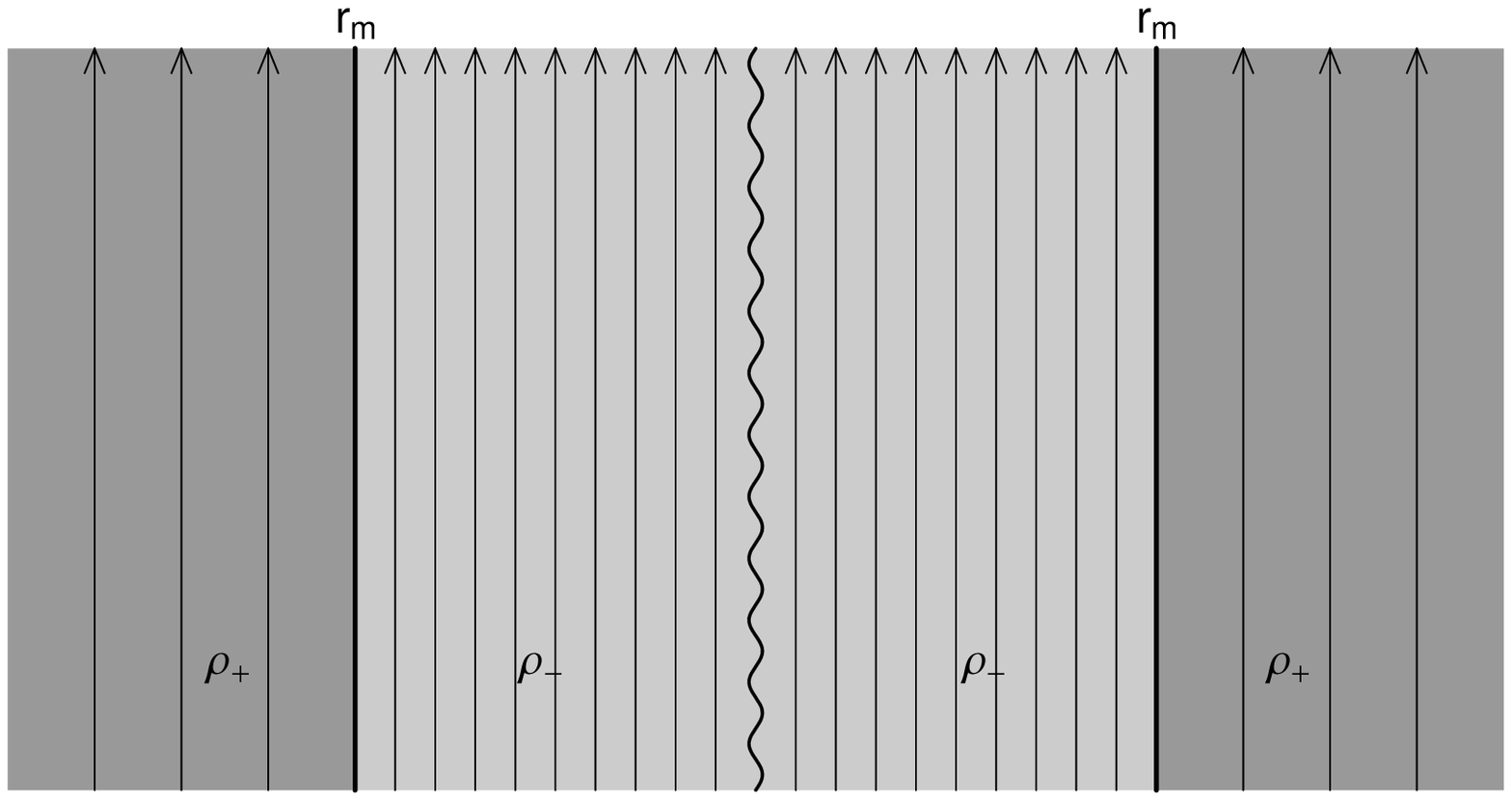}
\vspace{0.8cm}
\caption{Schematic representation of the $(r,z)$-section of the
cylindrical disk model studied in the paper. The wavy line represents
the central compact star. The two thick vertical lines represent the
radius $r = r_m$ which separates the magnetosphere on the inside from
the accretion disk on the outside.  The thin vertical lines with
arrows show magnetic field lines. In the region $r> r_m$, the mass
density ($\rho_+$) is high but the magnetic field is weak. In the
region $r< r_m$, the mass density ($\rho_-$) is low but the magnetic
field is strong. The jump in the mass density at $r = r_m$ is
described by the parameter $\mu$ defined in eq.~(\ref{mu}).
\label{fig1}}
\end{figure}

\clearpage
\begin{figure}
\epsscale{1}
\plotone{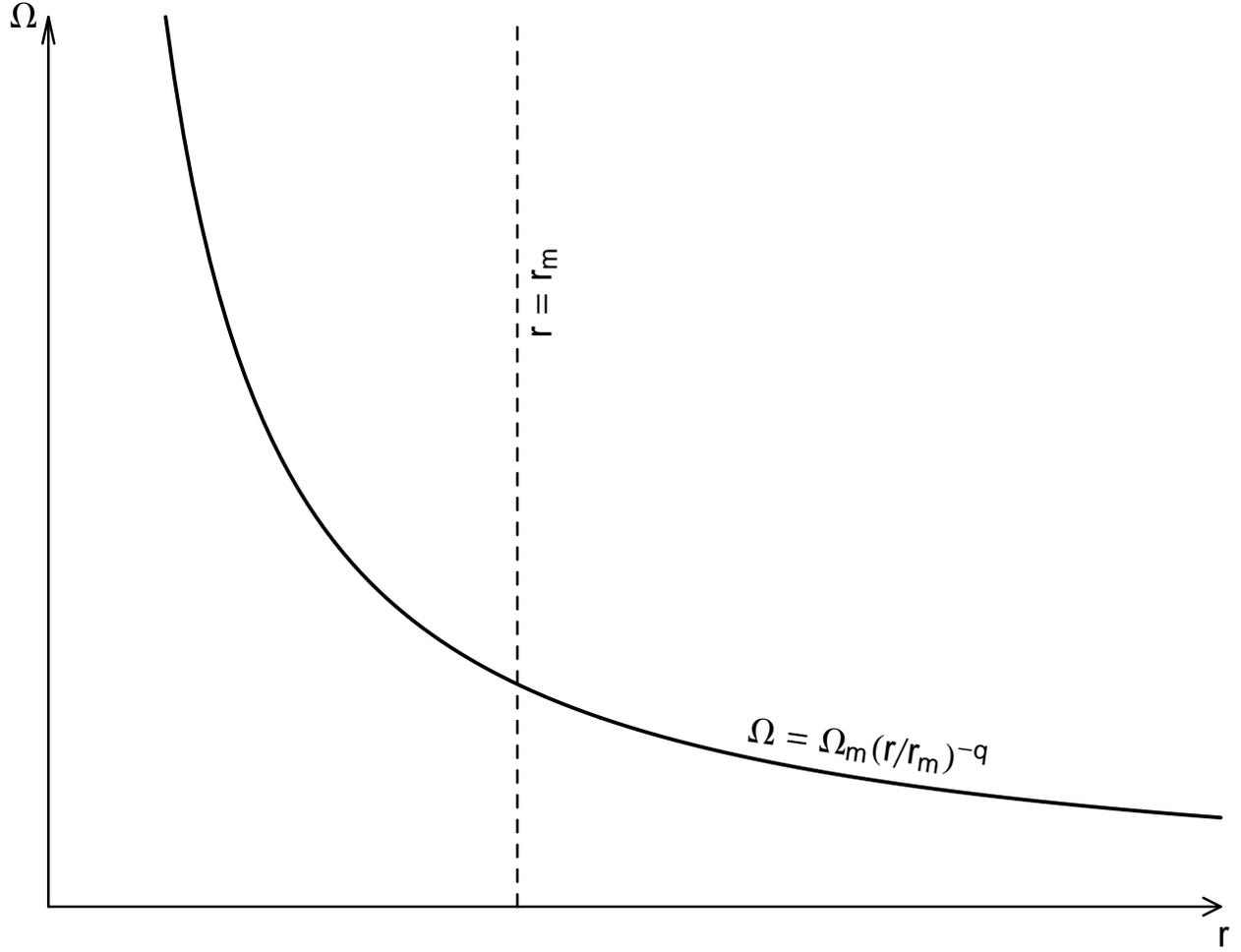}
\caption{Disk angular velocity as a function of radius for the case of
a central black hole (\S\ref{sec4.1}). The vertical dashed line marks
the magnetospheric radius $r_m$.  The angular velocity is taken to be
continuous across this boundary.
\label{fig2}}
\end{figure}

\clearpage
\begin{figure}
\epsscale{1}
\plotone{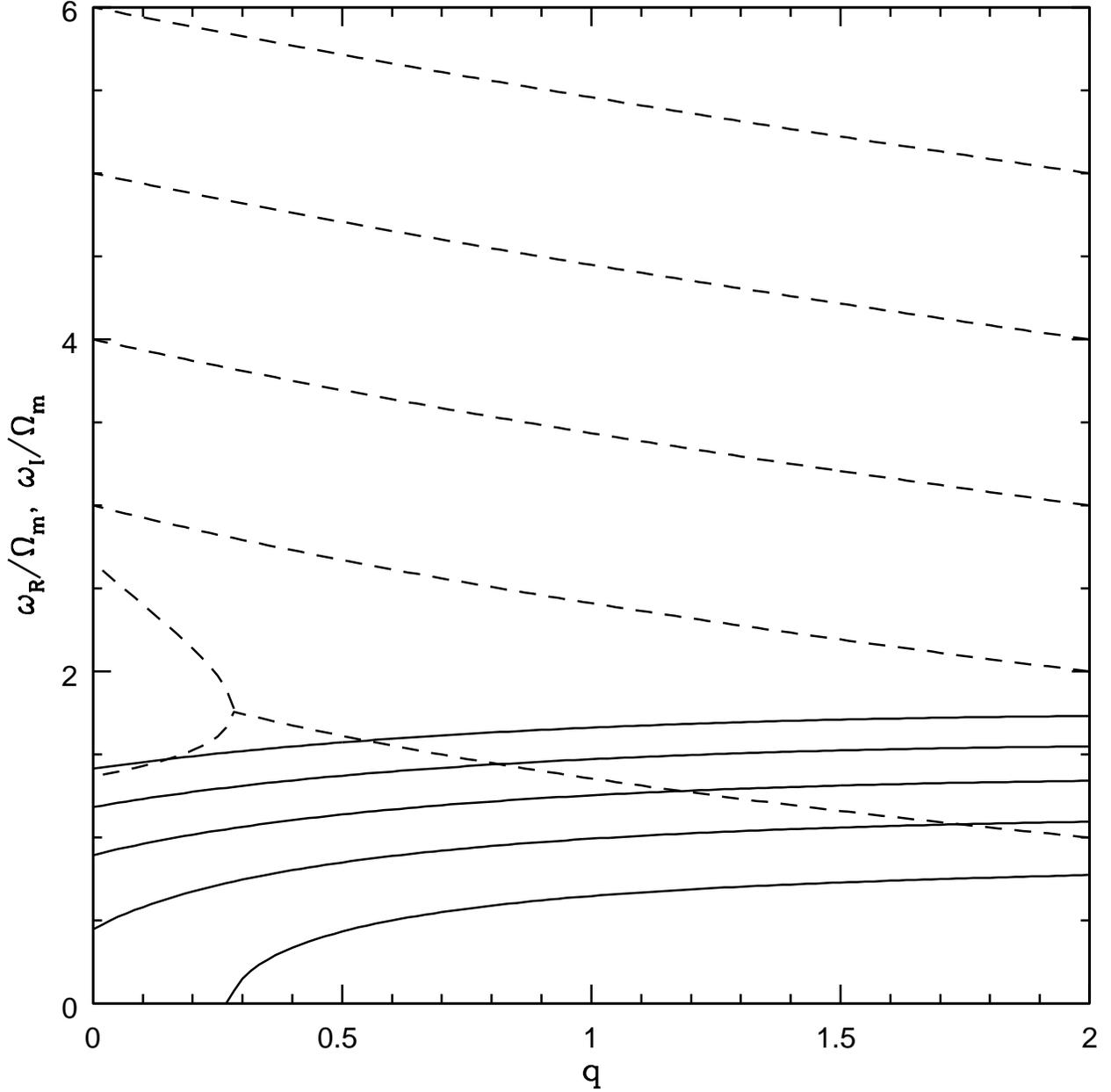}
\caption{Numerical solutions for the mode frequency as a function of
$q$ (defined by $\Omega\propto r^{-q}$). The mass density has a jump
at $r = r_m$ corresponding to $\mu=1$, and the density is assumed to
be constant on the two sides of the boundary. The effective gravity is
given by $\Omega_{{\rm eff}, m}^2 = 0.6 \Omega_m^2$. Solid lines
correspond to the imaginary part of the frequency and measure the
growth rates of the modes. Dashed lines correspond to the real part of
the frequency and represent the observed oscillation frequencies.
The different lines correspond to $m = 1$, $2$, $3$, $4$, and $5$
(bottom to top).
\label{fig3}}
\end{figure}

\clearpage
\begin{figure}
\epsscale{1}
\plotone{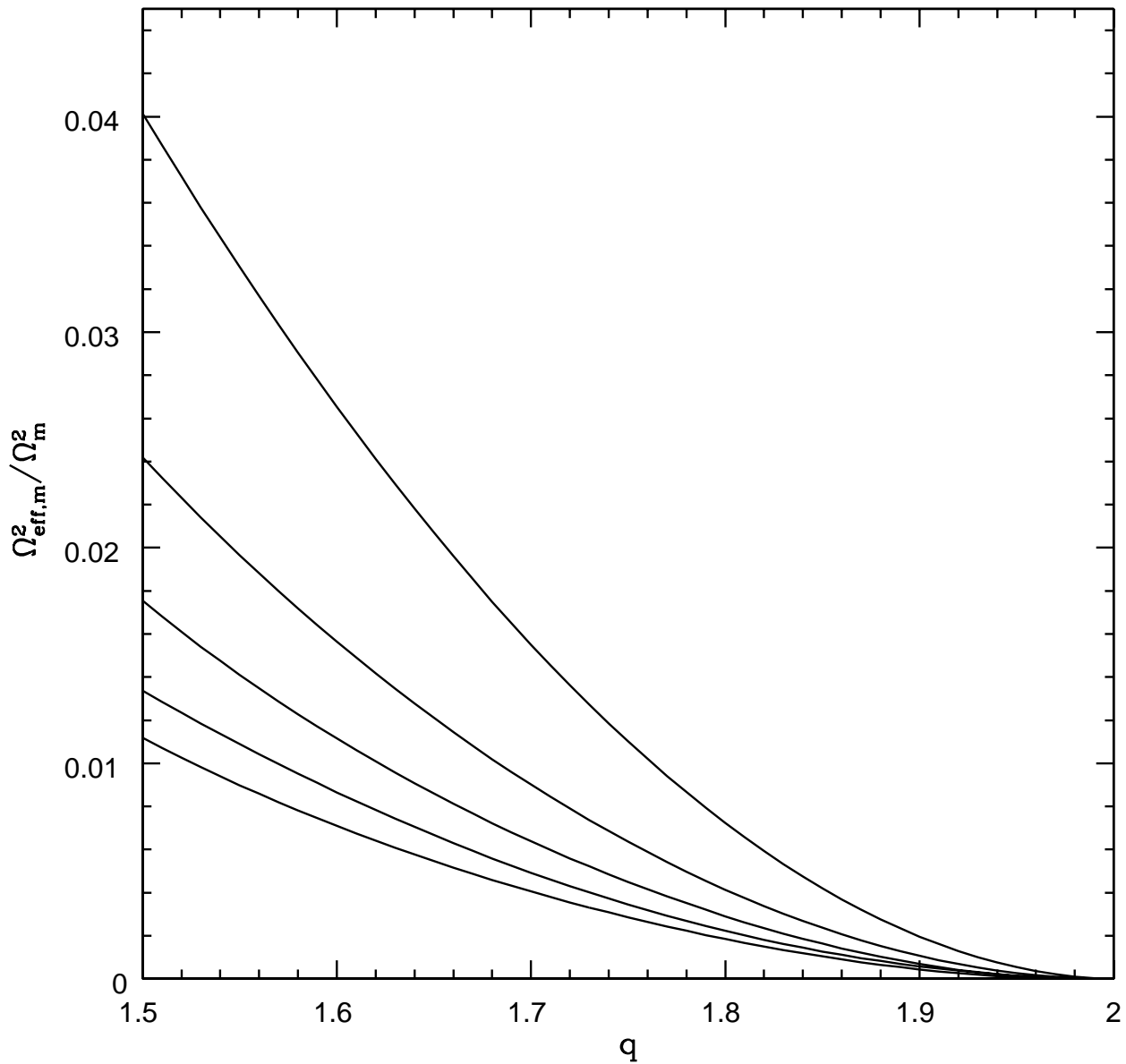}
\caption{Critical effective gravity frequency $\Omega_{{\rm
eff},m}^2$, which separates stable and unstable modes, for $3/2 \le q
\le 2$.  It is assumed that $\mu=1$ and $\gamma=0$.  The different
lines correspond to $m =1$, $2$, $3$, $4$, and $5$ (top to
bottom). Each line separates the $q$-$\Omega_{{\rm eff},m}^2$ plane
into two regions: above the line modes of the given $m$ are unstable,
and below the line the modes are stable.
\label{fig3a}}
\end{figure}

\clearpage
\begin{figure}
\epsscale{1}
\plotone{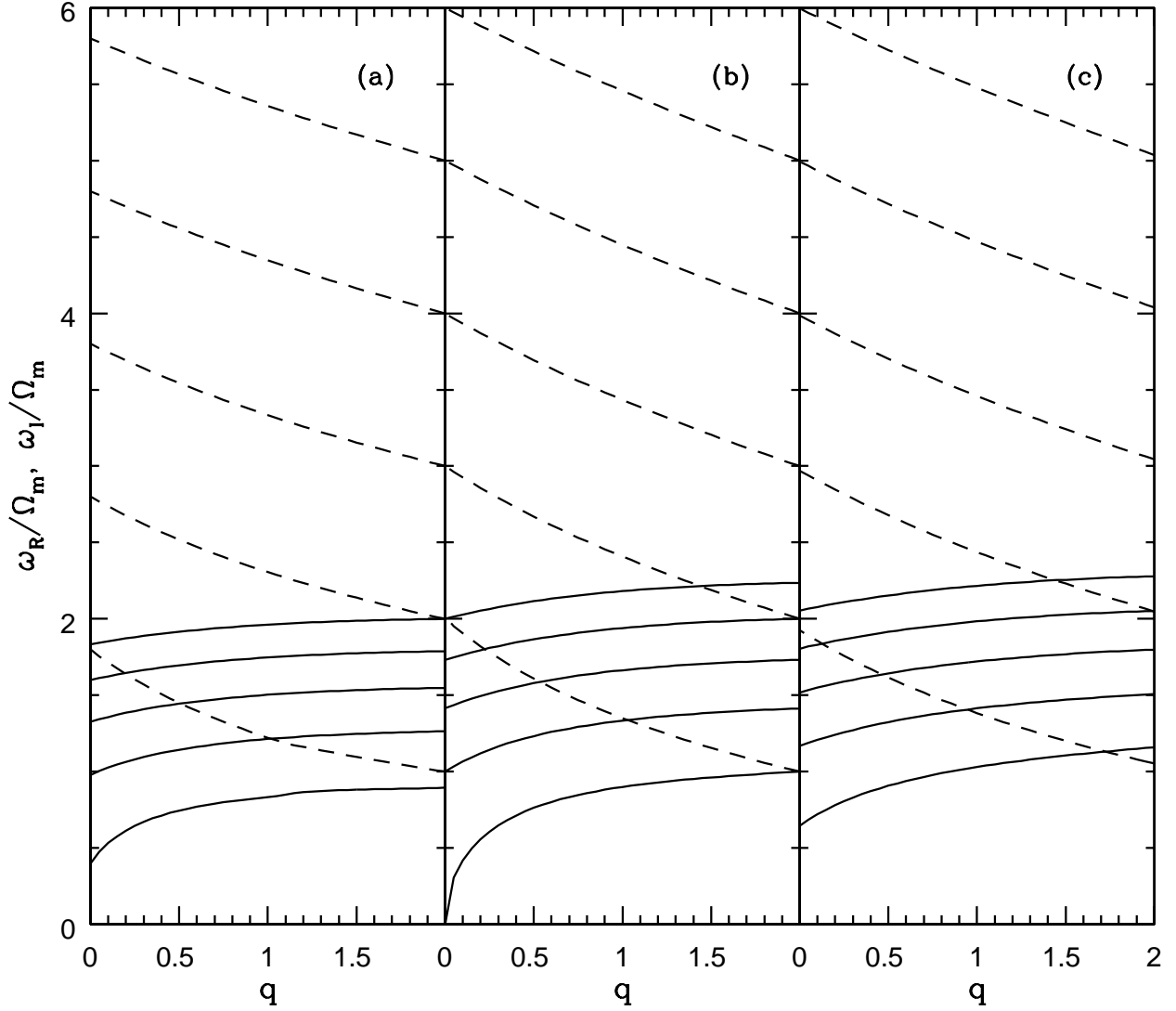}
\caption{Similar to Fig.~\ref{fig3}, but (a) $\mu = 0.8$,
$\Omega_{{\rm eff}, m}^2 = \Omega_m^2$, and the mass density is
constant on both sides of $r = r_{\rm s}$; (b) $\mu = 1$,
$\Omega_{{\rm eff},m}^2 =\Omega_m^2$, and the mass density is constant
on both sides of $r = r_m$; (c) $\mu = 1$, $\Omega_{{\rm eff},m}^2
=\Omega_m^2$, and the mass density is constant for $r < r_m$, but
decays with increasing radius according to $\rho\propto r^{-1}$ for $r
> r_m$.
\label{fig4}}
\end{figure}

\clearpage
\begin{figure}
\epsscale{1}
\plotone{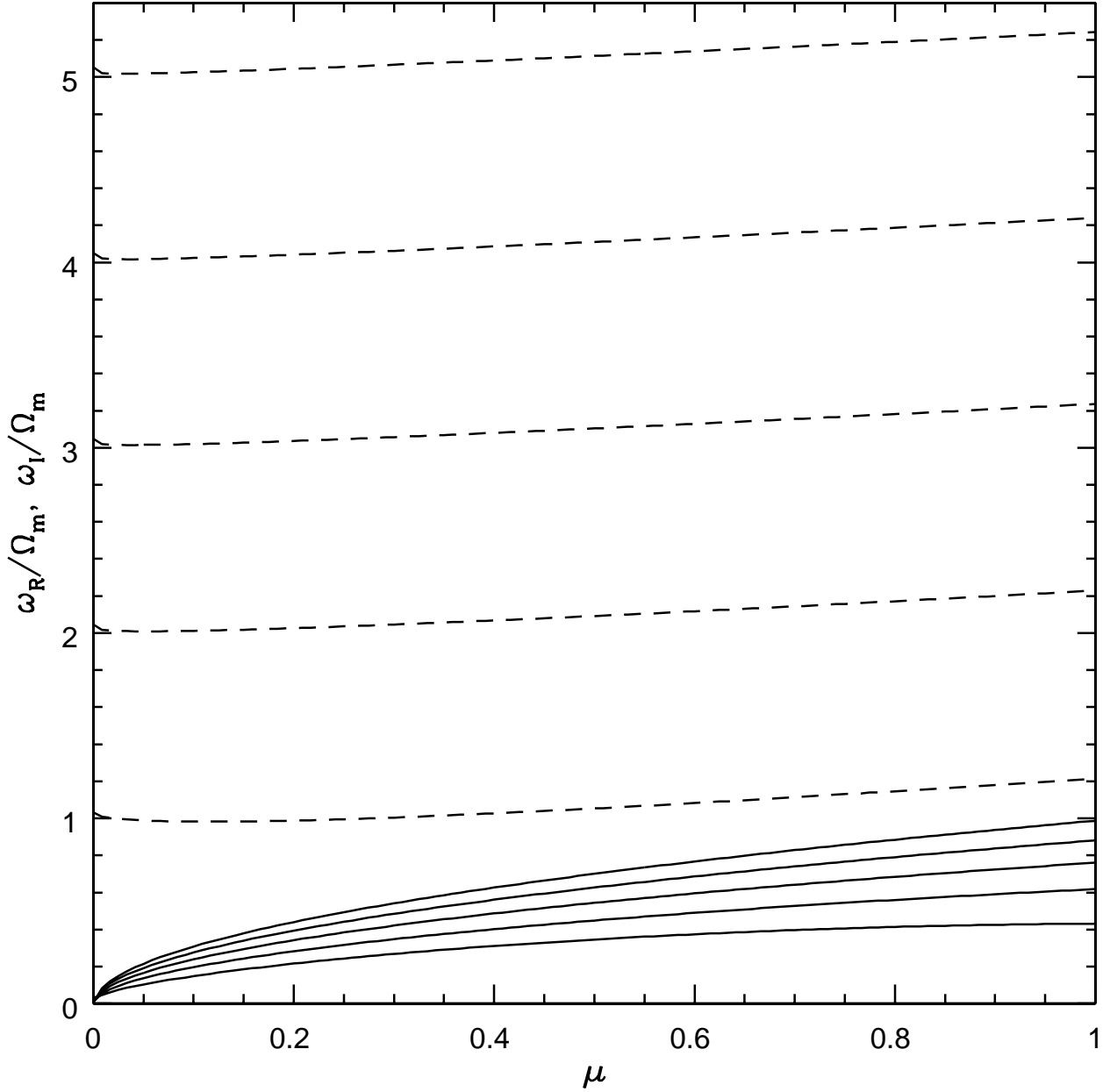}
\caption{Mode frequencies as a function of $\mu$ (defined by the jump
in mass density at $r = r_m$) for a Keplerian-type disk ($q =
3/2$). The mass density is constant for $r<r_{\rm m}$, but decays with
increasing radius according to $\rho\propto r^{-1}$ for $r > r_m$. At
$r = r_m$ the effective gravity frequency is $\Omega_{{\rm eff}, m}^2
= 0.2 \Omega_m^2$. Solid lines correspond to the imaginary part of the
frequency and dashed lines to the real part of the frequency.  The
different lines correspond to $m = 1$, $2$, $3$, $4$, and $5$ (bottom
to top).
\label{fig5}}
\end{figure}

\clearpage
\begin{figure}
\epsscale{1}
\plotone{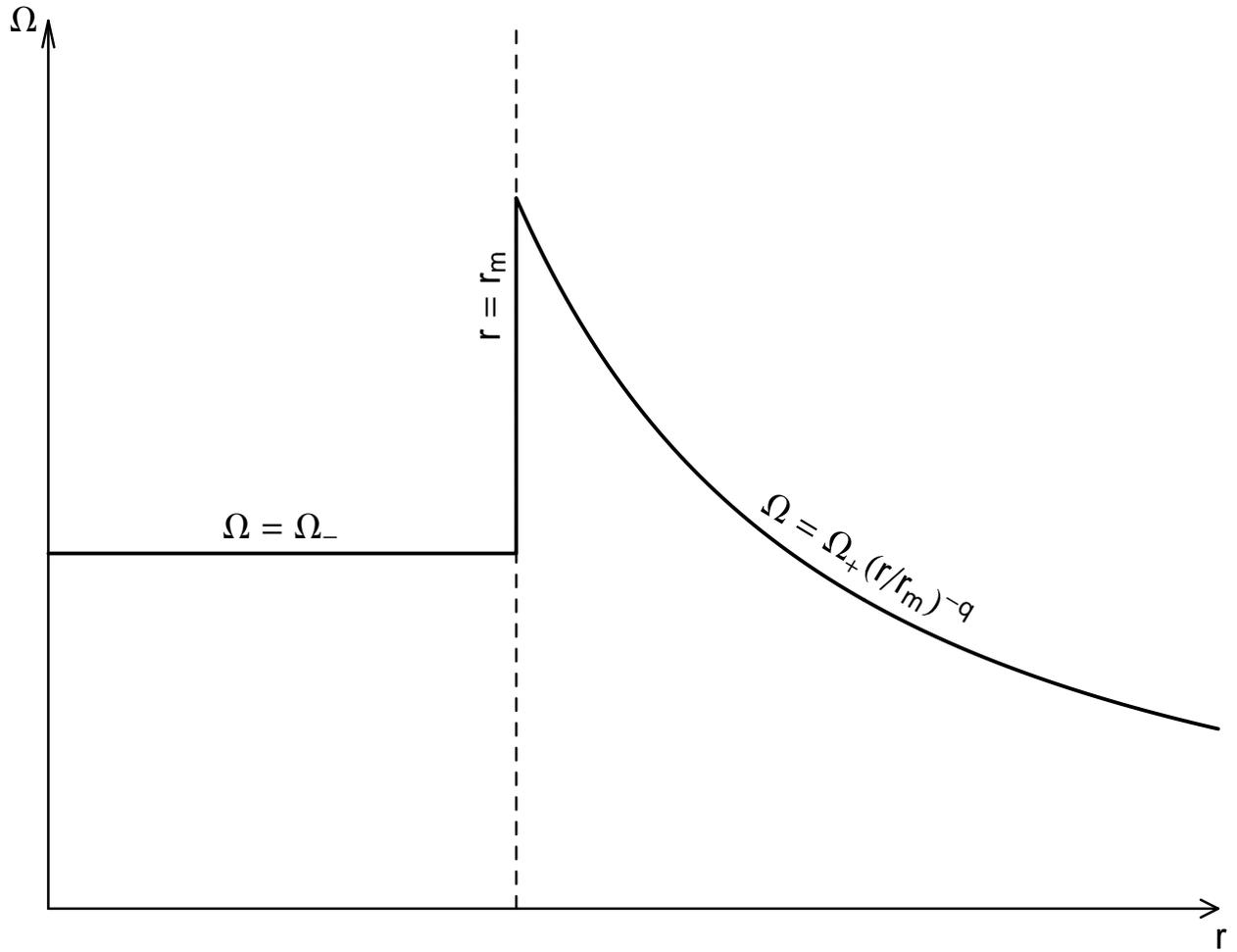}
\caption{Disk angular velocity as a function of radius for the case of
a magnetized neutron star (\S\ref{sec4.2}). The vertical dashed line
marks the magnetospheric radius $r_m$.  The angular velocity has a
jump at $r_m$: $\Delta_m (\Omega) = \Omega_+ - \Omega_- \neq 0$.
The jump may be either positive or negative.
\label{fig6}}
\end{figure}

\clearpage
\begin{figure}
\epsscale{1}
\plotone{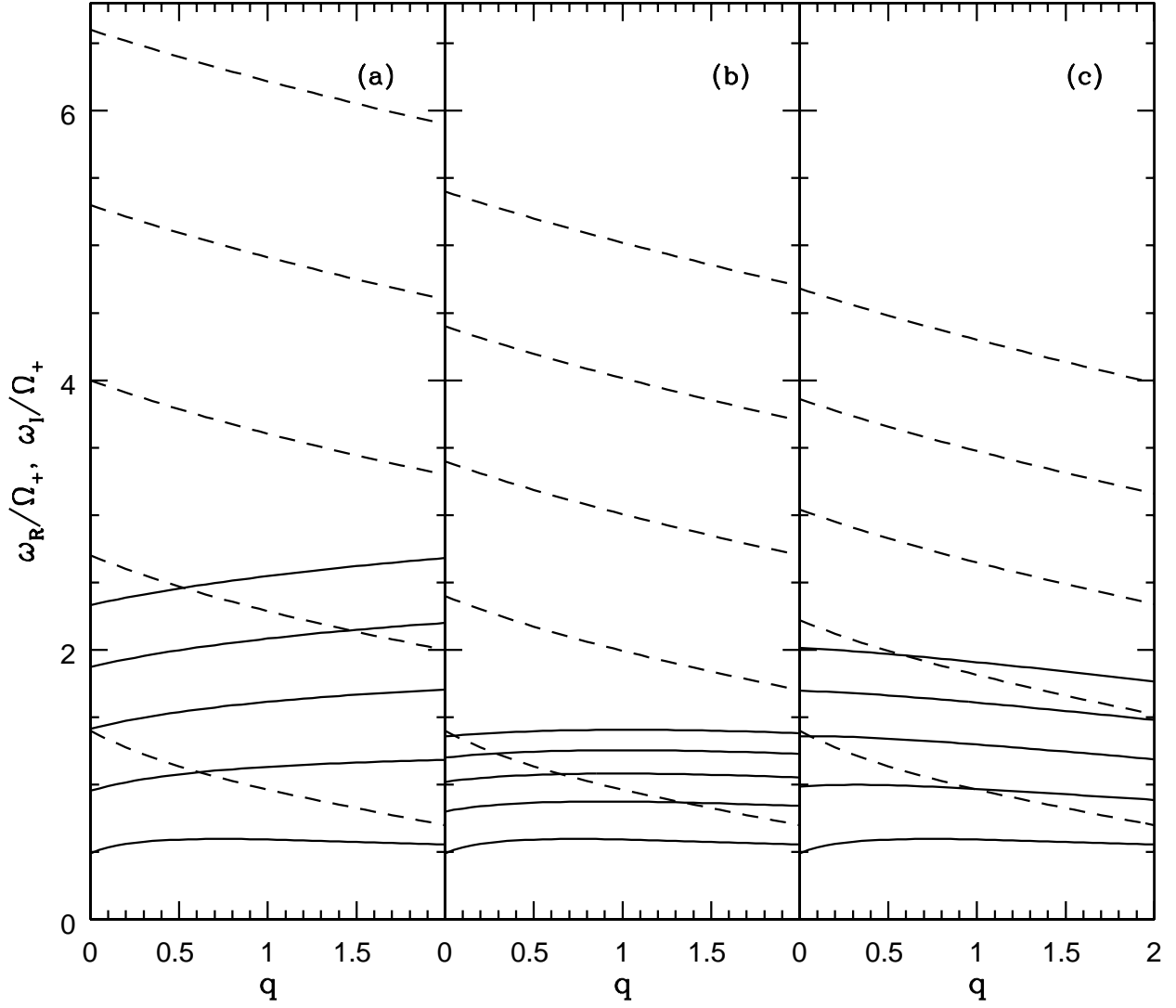}
\caption{Mode frequencies for a disk with an angular frequency jump
at $r_m$: $\Omega =\Omega_- = \mbox{constant}$ for $r< r_m$, and
$\Omega = \Omega_+ (r/r_m)^{-q}$ for $r>r_m$.  The mass density is
constant away from $r = r_m$, but has a jump at $r_m$ with $\mu =
0.4$. At $r = r_{m+}$, it is assumed that $\Omega_{\rm eff,+}^2 =
\Omega_+^2$. (a) The angular velocity is assumed to satisfy $\Omega_-
= 2\Omega_+$.  (b) $\Omega_- = \Omega_+$.  (c) $\Omega_- =
0.4\Omega_+$. Line types have the same meanings as in Fig.~\ref{fig3}.
\label{fig7}}
\end{figure}

\end{document}